\documentclass[12pt]{article}
\usepackage{fancyhdr}

\usepackage{mathrsfs}
\usepackage[T1]{fontenc}
\usepackage{mathpazo}
\usepackage{setspace}
\usepackage{amsfonts}
\usepackage{amssymb}
\usepackage{amsmath}
\usepackage{epsfig}
\usepackage{latexsym}
\usepackage{color}
\usepackage{graphicx}
\usepackage{nicefrac}
\usepackage[latin1]{inputenc}
\usepackage{slashed}
\usepackage{multirow}
\usepackage{comment}
\usepackage{soul}
\usepackage{hyperref}
\usepackage[nosort]{cite}
\usepackage{datetime}

\usepackage[titletoc]{appendix}

\def\hybrid{\topmargin -20pt    \oddsidemargin 0pt
        \headheight 0pt \headsep 0pt
        \textwidth 6.25in       
        \textheight 9.25in       
        \marginparwidth .875in
        \parskip 5pt plus 1pt   \jot = 1.5ex}

\hybrid

\def\baselinestretch{1.2}

\catcode`\@=11

\def\marginnote#1{}
\newcount\hour
\newcount\minute
\newtoks\amorpm
\hour=\time\divide\hour by60
\minute=\time{\multiply\hour by60 \global\advance\minute by-\hour}
\edef\standardtime{{\ifnum\hour<12 \global\amorpm={am}%
        \else\global\amorpm={pm}\advance\hour by-12 \fi
        \ifnum\hour=0 \hour=12 \fi
        \number\hour:\ifnum\minute<10 0\fi\number\minute\the\amorpm}}
\edef\militarytime{\number\hour:\ifnum\minute<10 0\fi\number\minute}

\def\draftlabel#1{{\@bsphack\if@filesw {\let\thepage\relax
   \xdef\@gtempa{\write\@auxout{\string
      \newlabel{#1}{{\@currentlabel}{\thepage}}}}}\@gtempa
   \if@nobreak \ifvmode\nobreak\fi\fi\fi\@esphack}
        \gdef\@eqnlabel{#1}}
\def\@eqnlabel{}
\def\@vacuum{}
\def\draftmarginnote#1{\marginpar{\raggedright\scriptsize\tt#1}}

\def\draft{\oddsidemargin -.5truein
        \def\@oddfoot{\sl preliminary draft \hfil
        \rm\thepage\hfil\sl\today\quad\militarytime}
        \let\@evenfoot\@oddfoot \overfullrule 3pt
        \let\label=\draftlabel
        \let\marginnote=\draftmarginnote
   \def\@eqnnum{(\theequation)\rlap{\kern\marginparsep\tt\@eqnlabel}%
\global\let\@eqnlabel\@vacuum}  }

\def\preprint{\twocolumn\sloppy\flushbottom\parindent 2em
        \leftmargini 2em\leftmarginv .5em\leftmarginvi .5em
        \oddsidemargin -.5in    \evensidemargin -.5in
        \columnsep .4in \footheight 0pt
        \textwidth 10.in        \topmargin  -.4in
        \headheight 12pt \topskip .4in
        \textheight 6.9in \footskip 0pt
        \def\@oddhead{\thepage\hfil\addtocounter{page}{1}\thepage}
        \let\@evenhead\@oddhead \def\@oddfoot{} \def\@evenfoot{} }

\def\numberbysection{\@addtoreset{equation}{section}
        \def\theequation{\thesection.\arabic{equation}}}

\def\underline#1{\relax\ifmmode\@@underline#1\else
        $\@@underline{\hbox{#1}}$\relax\fi}

\def\titlepage{\@restonecolfalse\if@twocolumn\@restonecoltrue\onecolumn
     \else \newpage \fi \thispagestyle{empty}\c@page\z@
        \def\thefootnote{\fnsymbol{footnote}} }

\def\endtitlepage{\if@restonecol\twocolumn \else \newpage \fi
        \def\thefootnote{\arabic{footnote}}
        \setcounter{footnote}{0}}  

\catcode`@=12
\relax

\def\figcap{\section*{Figure Captions\markboth
        {FIGURECAPTIONS}{FIGURECAPTIONS}}\list
        {Figure \arabic{enumi}:\hfill}{\settowidth\labelwidth{Figure
999:}
        \leftmargin\labelwidth
        \advance\leftmargin\labelsep\usecounter{enumi}}}
 \relax
\def\tablecap{\section*{Table Captions\markboth
        {TABLECAPTIONS}{TABLECAPTIONS}}\list
        {Table \arabic{enumi}:\hfill}{\settowidth\labelwidth{Table
999:}
        \leftmargin\labelwidth
        \advance\leftmargin\labelsep\usecounter{enumi}}}
 \relax
\def\reflist{\section*{References\markboth
        {REFLIST}{REFLIST}}\list
        {[\arabic{enumi}]\hfill}{\settowidth\labelwidth{[999]}
        \leftmargin\labelwidth
        \advance\leftmargin\labelsep\usecounter{enumi}}}
 \relax
\makeatletter
\newcounter{pubctr}
\def\publist{\@ifnextchar[{\@publist}{\@@publist}}
\def\@publist[#1]{\list
        {[\arabic{pubctr}]\hfill}{\settowidth\labelwidth{[999]}
        \leftmargin\labelwidth
        \advance\leftmargin\labelsep
        \@nmbrlisttrue\def\@listctr{pubctr}
        \setcounter{pubctr}{#1}\addtocounter{pubctr}{-1}}}
\def\@@publist{\list
        {[\arabic{pubctr}]\hfill}{\settowidth\labelwidth{[999]}
        \leftmargin\labelwidth
        \advance\leftmargin\labelsep
        \@nmbrlisttrue\def\@listctr{pubctr}}}
 \relax
\makeatother
\newskip\humongous \humongous=0pt plus 1000pt minus 1000pt

\newif\ifdtup

\relax

\def\be{\begin{equation}}
\def\ee{\end{equation}}
\def\ba{\begin{eqnarray}}
\def\ea{\end{eqnarray}}

\def\del{\partial}

\def\b{\beta}

\def\G{\Gamma}
\def\d{\delta}
\def\D{\Delta}

\def\th{\theta}

\def\m{\mu}

\def\l{\lambda}
\def\L{\Lambda}
\def\s{\sigma}

\def\cL{{\cal L}}

\def\no{\noindent}

\def\qq{\qquad}

\def\IR{\relax{\rm I\kern-.18em R}}

\def \J {{\bar J} }

\def \ha {{1\over 2}}

\def \ov {\over}

\def\IR{\relax{\rm I\kern-.18em R}}
\def\IL{\relax{\rm I\kern-.18em L}}

\def\inv{^{\raise.15ex\hbox{${\scriptscriptstyle -}$}\kern-.05em 1}}

\def\cL{{\cal L}}

\def\Tr{{\rm Tr}}

\begin{document}

\renewcommand{\theequation}{\thesection.\arabic{equation}}
\csname @addtoreset\endcsname{equation}{section}

\newcommand{\beq}{\begin{equation}}
\newcommand{\eeq}[1]{\label{#1}\end{equation}}
\newcommand{\ber}{\begin{equation}}
\newcommand{\eer}[1]{\label{#1}\end{equation}}
\newcommand{\eqn}[1]{(\ref{#1})}
\begin{titlepage}
\begin{center}


${}$
\vskip .2 in

{\large\bf Novel all loop actions of interacting CFTs:\\
 Construction, integrability and RG flows}

\vskip 0.4in

{\bf George Georgiou$^{1,2}$}\ and \ {\bf Konstantinos Sfetsos}$^{1}$
\vskip 0.15in

 {\em
${}^1$Department of Nuclear and Particle Physics,\\
Faculty of Physics, National and Kapodistrian University of Athens,\\
Athens 15784, Greece\\
}

\vskip 0.15in
{\em
${}^2$Institute of Nuclear and Particle Physics,
\\ National Center for Scientific Research Demokritos,
\\
Ag. Paraskevi, GR-15310 Athens, Greece
}

\vskip 0.12in

{\footnotesize \texttt georgiou@inp.demokritos.gr, ksfetsos@phys.uoa.gr}


\vskip .5in
\end{center}

\centerline{\bf Abstract}

\no
We construct the all loop effective action representing, for small couplings,
simultaneously self- and mutually interacting current algebra CFTs realized by WZW models.
This non-trivially generalizes our previous works where such interactions were, at the linear level, not simultaneously present.
For the two coupling case we prove integrability and calculate the coupled RG flow  
equations. We also consider non-Abelian T-duality type limits. Our models provide concrete realisations of  integrable flows between exact CFTs and exhibit 
several new features which we discuss in detail.

\vskip .4in
\noindent
\end{titlepage}
\vfill
\eject

\newpage

\tableofcontents

\noindent

\def\baselinestretch{1.2}
\baselineskip 20 pt
\noindent


\setcounter{equation}{0}
\section{Introduction }

The purpose of the present work is to derive the all-loop effective action and study important
properties such as integrability and the behavior under the renormalization group (RG) flow, 
of a class of theories based on WZW models for a group $G$ which encompasses and further
generalizes  all previous works in this research direction. 
Such models deviate from the conformal point in a way that can be quite involved. 
For small values of the coupling constants the deviation is driven by bilinears of the WZW model
chiral and anti-chiral currents, denoted by $J^a_\pm$, where $a=1,2,\dots,\dim G$. These perturbations drive the theory away from the conformal point since they are generically not
exactly marginal. Their form may serve also  to distinguish between the different type of models
existing in the literature and also singles out the present work in comparison with previous ones. 

\no
The first such 
example was worked out in \cite{Sfetsos:2013wia} where the unperturbed conformal 
field theory (CFT) was a single WZW model for a group $G$ and level $k$ for the corresponding 
current algebra. In this case the perturbation contains terms proportional to
\be
\label{selff}
 J^a_+ J^b_-\ .
\ee
The currents belong to the same original CFT, so that this is a theory 
of self-interactions. 

\no
The next natural step in this program was to consider the case in which the original CFT is composed by two WZW models with currents $J^a_{i \pm}$, with $i=1,2$ and 
the perturbation being a linear combination of the two current bilinears of the form
\be
\label{mutuall}
 J^a_{1+} J^b_{2-}\ ,\qquad  J^a_{2+} J^b_{1-}\ .
\ee
Such perturbations presents a mutual interaction of the two WZW model theories and there are no self-interaction terms of currents in the same WZW model.
The effective action for this case was constructed for equal current algebra levels in \cite{Georgiou:2016urf} and extended to the unequal level case in \cite{Georgiou:2017jfi}
which, although technically similar to the equal level case, encodes some new important physical features.
An extension of these with several WZW models mutually interacting was constructed in \cite{Georgiou:2017oly}.

\no
These models were generically called $\l$-deformed from the letter-symbol $\l_{ab}$ 
used for the coefficients multiplying the perturbative terms in  \eqn{selff} and \eqn{mutuall}. 
Beyond the linear level the effective actions  \cite{Sfetsos:2013wia} and \cite{Georgiou:2016urf,Georgiou:2017jfi} have of course non-trivial dependence on these $\l$-parameters. 

\no
Although the $\s$-model all-loop effective actions 
corresponding to the perturbations \eqn{selff} and \eqn{mutuall} are different, certain quantum characteristics of the models, concerning in particular the renormalization group (RG) are 
the same or closely related.  
To explain that, let us first recall that the RG flow equations for the $\s$-model of \cite{Sfetsos:2013wia} were computed, exactly in $\l$ and for large 
$k$, using gravitational methods \cite{Itsios:2014lca,Sfetsos:2014jfa} with results in perfect 
agreement  with those obtained in the past using field theoretical  methods \cite{Kutasov:1989dt, Gerganov:2000mt,LeClair:2001yp} and more recently in \cite{Appadu:2015nfa}. 
In addition, all-loop correlators of current and primary field operators have been computed in 
\cite{Georgiou:2015nka,Georgiou:2016iom,Georgiou:2016zyo}. In these computations a few terms obtained using perturbation theory and the non-perturbative symmetry, argued via path integral considerations in \cite{Kutasov:1989aw}, were enough to obtain the exact results.
In addition, for the $\s$-model of \cite{Georgiou:2016urf,Georgiou:2017jfi}
the anomalous dimensions of current and primaries in this theory were computed using CFT techniques in a field theoretical approach and symmetry arguments in \cite{Georgiou:2017aei}.
As it turns out the $\b$-functions for the couplings are identical to those of several non-coupled 
single $\l$-deformed models. Also,
the anomalous dimensions of
currents are related, though for the anomalous dimensions of generic primary field operators the results differ.
The reason for such remarkable agreements is the fact that from a CFT point of view 
each of the two terms in \eqn{mutuall} is the same as the one in \eqn{selff} and moreover these
two terms have vanishing operator product expansion (OPE), i.e. are mutually non-interacting. Hence, the corresponding couplings constants run independently under the RG flow and similar arguments can be made for the anomalous dimensions. In further support of the above, 
one can show \cite{Georgiou:2017oly} that the effective action of \cite{Georgiou:2016urf} is canonically equivalent to the sum of two actions as in \cite{Sfetsos:2013wia}.

\no
We note that the most general
expressions for the $\beta$-functions and anomalous dimensions for the operator driving 
the perturbation can be found in \cite{Sagkrioti:2018rwg}. 
This includes the
most general coupling matrix $\l_{ab}$ and having different levels for the chiral and anti-chiral 
currents. 
Remarkably, the above developments allowed the computation of Zamolodchikov's $C$-function \cite{Zamolodchikov:1986gt} exactly in the deformation 
parameter for the case of isotropic perturbations and to leading order in $k$  in \cite{c-function:2018}. The similarities mentioned above 
for the models corresponding to \eqn{selff} and \eqn{mutuall} extend to this case as well,
the reason being the close relation of the $\b$- and $C$-functions \cite{Zamolodchikov:1986gt}.

\no
There were many other parallel developments or closely related to the above.  
Of particular importance is the extension to cases
where the unperturbed CFT of a single WZW model is replaced by a coset CFT \cite{Sfetsos:2013wia,Hollowood:2014rla,Hollowood:2014qma,Sfetsos:2017sep}.
The corresponding analysis for the case of supergroups was considered in \cite{Hollowood:2014rla,Hollowood:2014qma}. 
In addition, though integrability has not been a key factor in the computation 
of the $\b$-functions
and of the operators anomalous dimensions, in the case of isotropic deformations the above models have been demonstrated to be integrable \cite{Sfetsos:2013wia,Hollowood:2014rla,Hollowood:2014qma,Itsios:2014vfa}, \cite{Sfetsos:2017sep} and 
\cite{Georgiou:2016urf,Georgiou:2017jfi}. For the particular case of the isotropic deformation based on $SU(2)$ this has been proven before in \cite{Balog:1993es}.
Integrability
was shown to persist in some other cases with more deformation parameters \cite{Sfetsos:2014lla,Sfetsos:2015nya}.
Furthermore, deformed models of low dimensionality have been embedded to supergravity \cite{Sfetsos:2014cea,Demulder:2015lva,Borsato:2016zcf,Chervonyi:2016ajp}.
Moreover, $\l$-deformations are related via Poisson-Lie T-duality,
introduced for group spaces in  \cite{KS95a} and extended for coset spaces in 
\cite{Sfetsos:1999zm},
 and appropriate analytic continuations  \cite{Vicedo:2015pna,Hoare:2015gda}, \cite{Sfetsos:2015nya,Klimcik:2015gba,Klimcik:2016rov,Hoare:2018ebg} to $\eta$-deformations for group and coset spaces which were introduced in \cite{Klimcik:2002zj,Klimcik:2008eq,Klimcik:2014}
and \cite{Delduc:2013fga,Delduc:2013qra,Arutyunov:2013ega}, respectively.
The dynamics of scalar fields in some $\l$-deformed geometries corresponding to 
coset CFTs has been discussed in \cite{Lunin:2018vsn}, the relation to Chern-Simons theories
in \cite{Schmidtt:2017ngw} and  D-branes in the context of $\lambda$-deformations in 
\cite{Driezen:2018glg}.

\no
A very important remaining question is to construct a theory in which all current bilinears constructed form 
the original CFT based on two WZW models play r\^ole in the perturbation, that is all 
terms of the following type are present at the linear level and on equal footing
\be
\label{alll}
J^a_{1+} J^b_{1-}\ ,\qquad J^a_{2+} J^b_{2-}\ ,\qquad
J^a_{1+} J^b_{2-}\ ,\qquad J^a_{2+} J^b_{1-}\ .
\ee
In this paper we overtake precisely this task and construct in section 2
the effective action taking into account all
loop effects corresponding to the simultaneous presence of all of the above perturbations, 
self as well as mutual. In this case all terms have non-vanishing OPEs with each other at a sufficiently high order in perturbation theory. 
Therefore it its expected that the $\b$-functions and anomalous dimensions for the operators
will  generically depend on all coupling constants. 
We focus for simplicity to a particular two parameter model case. We will provide a proof that the model is integrable and we will construct a non-Abelian T-duality limit in section 3. In section 4 we will  derive and study in 
detail the RG flow equations for these couplings. We conclude the paper in section 5.

\section{The Lagrangian and the equations of motion}

In this section we construct our effective actions and the corresponding equations of motion.

Consider the group elements $g_i$, $i=1,2$ in a group $G$ and the corresponding 
actions for  two WZW models at levels $k_1$ and $k_2$. 
We add to them the action of two PCMs which are 
mutually interacting and are constructed by two group elements $\tilde g_i$, $i=1,2$
 in the same group $G$. Namely, we have that
\be
\label{oractt}
\begin{split}
& S_{k_i,E_i}(g_i,\tilde g_i)= S_{k_1}(g_1)+ S_{k_2}(g_2)
-{1\ov \pi}\int d^2\s\ \Big( \tilde g_1^{-1}\del_+\tilde g_1  E_1 \tilde g_1^{-1} \del_-\tilde g_1
\\
&\qq\qq
+ \tilde g_1^{-1}\del_+\tilde g_1  E_2 \tilde g_2^{-1} \del_-\tilde g_2 + 
\tilde g_2^{-1}\del_+\tilde g_2  E_3 \tilde g_1^{-1} \del_-\tilde g_1
+ \tilde g_2^{-1}\del_+\tilde g_2  E_4 \tilde g_2^{-1} \del_-\tilde g_2 \Big)\  ,
\end{split}
\ee
where the $E_i$, $i=1,2,3,4$ are generic coupling matrices. 
In the spirit of \cite{Sfetsos:2013wia,Georgiou:2016urf} we gauge the global 
symmetry acting on the group elements as 
$g_i\to \L_i^{-1} g_i\L_i$ and $\tilde g_i\to \L_i^{-1} \tilde g_i$, $i=1,2$. Hence, we will consider the 
gauge invariant action 
\ba
\label{gauacc}
&& S_{k_i,E_i}(g_i,\tilde g_i, A_\pm, B_\pm)  = S_{k_1}(g_1,A_\pm) + S_{k_2}(g_2,B_\pm) 
-{1\ov \pi}\int d^2\s\  \Big( \tilde g_1^{-1}D_+\tilde g_1  E_1 \tilde g_1^{-1} D_-\tilde g_1
\nonumber\\
&&
 \qq \ + \tilde g_1^{-1}D_+\tilde g_1  E_2 \tilde g_2^{-1} D_-\tilde g_2 + 
\tilde g_2^{-1}D_+\tilde g_2  E_3 \tilde g_1^{-1} D_-\tilde g_1
+ \tilde g_2^{-1}D_+\tilde g_2  E_4 \tilde g_2^{-1} D_-\tilde g_2 \Big)\ ,
\ea
where the standard gauged WZW action is
\be
\begin{split}
&  S_{k_1}(g_1,A_\pm) = S_{k_1}(g_1)
+{k_1\ov \pi} \int d^2\s \ \Tr \big(A_- \del_+ g_1 g_1^{-1}   - A_+ g_1^{-1} \del_- g_1
\\
& \qq\qq\qq + A_- g_1 A_+ g_1^{-1}-A_-A_+\big)
\end{split}
\ee
and similarly for  $ S_{k_2}(g_2,B_\pm)$. The covariant derivatives are defined as
$D_\pm \tilde g_1= (\del_\pm -A_\pm) \tilde g_1$ 
and $D_\pm \tilde g_2= (\del_\pm -B_\pm) \tilde g_2$.
After fixing the gauge in \eqn{gauacc} as $\tilde g_1=\tilde g_2=\mathbb{1}$ we arrive at the
following action
\be
\begin{split}
&  S_{k_i,\l_i}(g_1,g_2, A_\pm, B_\pm) = S_{k_1}(g_1) + S_{k_2}(g_2)
\\
&\qq\quad +{k_1\ov \pi} \int d^2\s \ \Tr \big(A_- \del_+ g_1 g_1^{-1}   - A_+ g_1^{-1} \del_- g_1+ A_- g_1 A_+ g_1^{-1}\big) \\
& \qq\quad  +{k_2\ov \pi} \int d^2\s \ \Tr \big(B_- \del_+ g_2 g_2^{-1}   - B_+ g_2^{-1} \del_- g_2+ B_- g_2 B_+ g_2^{-1} \big)
\\
&\qq\quad
- {\sqrt{k_1 k_2}\ov \pi} \int d^2\s\ \Big( B_+ \l_1^{-1} A_-  +A_+ \l_2^{-1} B_-
+ B_+ \l_3^{-1} B_-  +A_+ \l_4^{-1} A_-\Big)\ ,
\label{gaufix}
 \end{split}
\ee
where for later convenience 
we have redefined the coupling matrices appearing in the PCM models as
\be
\begin{split}
& \sqrt{k_1k_2}\ \l_1^{-1}= E_3 \ ,\qq \sqrt{k_1k_2}\ \l_2^{-1}= E_2\ ,
\\
&
 \sqrt{k_1k_2}\ (\l_3^{-1}-\l_0^{-1})= E_4 \ ,\qq \sqrt{k_1k_2}\ (\l_4^{-1}-\l_0)= E_1\ .
\end{split}
\ee 
In order to obtain the $\s$-model we can integrate out the gauge fields since they appear 
only quadratically. To do that we use their equations of motion which we prefer to present 
later  in \eqn{dggd}. In this way we find that
\be
\label{abp}
\begin{split}
& A_+  =  i \Big((\l_0 \l_3^{-T} - D_2)\l_1^T (\l_0^{-1} \l_4^{-T} - D_1)-\l_2^{-T}\Big)^{-1}
\Big((\l_0 \l_3^{-T} - D_2)\l_1^T J_{1+} -\l_0^{-1} J_{2+}\Big)\ ,
\\
&
B_+ =  i \Big((\l_0^{-1} \l_4^{-T} - D_1)\l_2^T (\l_0 \l_3^{-T} - D_2)-\l_1^{-T}\Big)^{-1}
\Big((\l_0^{-1} \l_4^{-T} - D_1)\l_2^T J_{2+} -\l_0 J_{1+}\Big)
\end{split}
\ee
and that
\be
\label{abm}
\begin{split}
& A_- =- i \Big((\l_0 \l_3^{-1} - D_2^T)\l_2 (\l_0^{-1} \l_4^{-1} - D_1^T)-\l_1^{-1}\Big)^{-1}
\Big((\l_0 \l_3^{-1} - D_2^T)\l_2 J_{1-} - \l_0^{-1} J_{2-} \Big)\ ,
\\
&
B_- = - i \Big((\l_0^{-1} \l_4^{-1} - D_1^T)\l_1 (\l_0 \l_3^{-1} - D_2^T)-\l_2^{-1}\Big)^{-1}
\Big((\l_0^{-1} \l_4^{-1} - D_1^T)\l_1 J_{2-} -\l_0 J_{1-}\Big)\ .
\end{split}
\ee
The definition of the matrices $D_{ab}$ and the currents $J^a_{\pm}$ is  as follows
\be
\label{hg3}
J^a_+ = - i\, \Tr(t^a \del_+ g g^{-1}) ,\qq J^a_- = - i\, \Tr(t^a g^{-1}\del_- g )\ ,
\qq D_{ab}= \Tr(t_a g t_b g^{-1})\  ,
\ee
where the $t^a$'s are Hermitian matrices obeying $[t_a,t_b]=i f_{ab}{}^c t_c$, for some
real algebra structure constants. 
When a current or the orthogonal matrix $D$ has an index $1$ or $2$ this implies 
that one should use the corresponding group element in its definition.
In addition, we have defined the ratio of the two levels
\be
\label{levl0}
 \l_0=\sqrt{{k_1 \ov k_2}}\ .
\ee

\no
Substitution of the expressions for the gauge fields into \eqn{gaufix} results into a
$\s$-model action which can be written in matrix notation as
\be
\begin{split}
&  S_{k_i,\l_i}(g_1,g_2) = S_{k_1}(g_1) + S_{k_2}(g_2)
\\
& + {1\ov \pi} \int  d^2\s\ \Big[  k_1 J_{1+}
\Big((\l_0 \l_3^{-1} - D_2^T)\l_2 (\l_0^{-1} \l_4^{-1} - D_1^T)-\l_1^{-1}\Big)^{-1}
\Big((\l_0 \l_3^{-1} - D_2^T)\l_2 J_{1-} - \l_0^{-1} J_{2-} \Big)
\\
&\quad
+ k_2 J_{2+} \Big((\l_0^{-1} \l_4^{-1} - D_1^T)\l_1 (\l_0 \l_3^{-1} - D_2^T)-\l_2^{-1}\Big)^{-1}
\Big((\l_0^{-1} \l_4^{-1} - D_1^T)\l_1 J_{2-} -\l_0 J_{1-}\Big)\Big ]\ .
\end{split}
\label{defactigen}
\ee
Note that, for $\l_{1,2}\to \infty$ one obtains two decoupled single $\l$-deformed models 
\cite{Sfetsos:2013wia}. These can be most easily seen by taking this limit in \eqn{gaufix} before 
integrating out the gauge fields.  Hence, then \eqn{defactigen} describes  
self-interactions for two decoupled WZW models  
which for small values of the remaining couplings $\l_{3,4}$ are of the form \eqn{selff}. 
On the other hand if $\l_{3,4}\to \infty$ one obtains the model of \cite{Georgiou:2016urf} corresponding for small values of $\l_{1,2}$ to mutual interactions of two WZW models of
the form \eqn{mutuall}.
This may also easily seen from inspecting \eqn{gaufix}.

By keeping all matrices one has the most general scenario. Indeed, if we take small values
for the entries of the $\l$-matrices keeping nevertheless their ratios finite, we obtain that
\be
\label{milit}
\begin{split}
&  S_{k_i,\l_i}(g_1,g_2) = S_{k_1}(g_1) + S_{k_2}(g_2)
 + {1\ov \pi} \int  d^2\s\ \Big[ k_1 \l_0 M_2  J_{1+} J_{1-} 
 \\
&\qq 
- k_1 \l_0^{-1} M_3  J_{1+} J_{2-}
+  k_2 \l_0^{-1} M_1  J_{2+} J_{2-} - k_1 \l_0 M_4  J_{2+} J_{1-}\Big] + \cdots \ ,
\end{split}
\ee
where we took for simplicity all the $\l$-matrices proportional to the identity and we have defined 
the constants 
\be
(M_1,M_2,M_3,M_4) = {1\ov \l_1 \l_2 -\l_3 \l_4} (\l_1\l_2\l_3,\l_1\l_2\l_4,\l_1\l_3\l_4,\l_2\l_3\l_4)
\ .
\ee
These constant couplings are small and of the same order of magnitude as the original $\l$'s.
Hence, what drives the original CFT, which is given by the sum of 
the two WZW actions in \eqn{milit}, away from the conformal point is indeed a linear combination 
of all terms in \eqn{alll} representing  simultaneously self- and mutual-interactions.

We will write in some detail the equations of motion since this will be convenient in demonstrating that the theory described by 
the actions \eqn{gaufix} and \eqn{defactigen} is 
integrable for a particular case where two out of the four couplings are present.

\no
Varying \eqn{gaufix} with respect
to $A_\pm$ and $B_\pm$ we find the following constraints
\be
\label{dggd}
\begin{split}
&
D_+ g_1\, g_1^{-1} =  (\l_0^{-1}\l_4^{-T}-\mathbb{1}) A_+ + \l_0^{-1} \l_1^{-T} B_+ \ ,
\\
&
g_1^{-1} D_- g_1 = (\mathbb{1}-\l_0^{-1}\l_4^{-1}) A_- - \l_0^{-1} \l_2^{-1} B_-\ ,
\\
&
D_+ g_2\, g_2^{-1} =  (\l_0\l_3^{-T}-\mathbb{1}) B_+ + \l_0 \l_2^{-T} A_+\ ,
\\
&
g_2^{-1} D_- g_2 =  (\mathbb{1}-\l_0 \l_3^{-1}) B_- - \l_0 \l_1^{-1} A_- \ ,
\end{split}
\ee
where the covariant derivatives acting on the group elements are defined according to the transformation laws that leave \eqn{gauacc} invariant. Namely,
$D_\pm g_1= \del_\pm g_1 -[A_\pm,g_1]$ and $D_\pm g_2= \del_\pm g_2 -[B_\pm,g_2]$.
By solving these for the gauge fields we obtain the solution \eqn{abp} and \eqn{abm} we have already presented.
Varying the action with respect to group elements $g_1$ and $g_2$ results into
\be
\label{eqg1g2}
D_ -(D_+ g_1 g_1^{-1})= F_{+-}^{(A)}\ ,\qq D_ -(D_+ g_2 g_2^{-1})
= F_{+-}^{(B)}\ ,
\ee
where the field strenghts are defined as usual
\be
F_{+-}^{(A)}=\del_+ A_- - \del_- A_+ - [A_+,A_-]\ ,\qq
F_{+-}^{(B)}=\del_+ B_- - \del_- B_+ - [B_+,B_-]\ .
\ee
Equivalently, the equations \eqref{eqg1g2} can be written as
\be
D_+(g_1^{-1}D_- g_1)= F_{+-}^{(A)}\ ,\qq
D_+(g_2^{-1}D_- g_2)= F_{+-}^{(B)}\ .
\label{eqg1g22}
\ee
The next step is to substitute the constraint equations \eqn{dggd} in \eqn{eqg1g2} and \eqn{eqg1g22}. After some algebra one obtains the following
two sets of equations
\be
\begin{split}
\label{eomAinitial1}
&\del_+ A_- - \l_0^{-1}\l_4^{-T} \del_-  A_+ - \l_0^{-1}\l_1^{-T} \del_-B_+
= \l_0^{-1} [\l_4^{-T} A_+,A_-] +\l_0^{-1} [\l_1^{-T} B_+,A_-]\ ,
\\
& \del_+ B_-  - \l_0\l_3^{-T}\del_-  B_+ - \l_0 \l_2^{-T} \del_- A_+ =
\l_0 [\l_3^{-T} B_+,B_-] + \l_0 [\l_2^{-T} A_+,B_-]\
\end{split}
\ee
and 
\be
\begin{split}
\label{eomAinitial2}
&  
\l_0^{-1}\l_2^{-1}\del_+ B_- -  \del_-A_+  +  \l_0^{-1} \l_4^{-1} \del_+ A_-
= \l_0^{-1} [A_+,\l_4^{-1} A_-] + \l_0^{-1} [A_+,\l_2^{-1}B_-]  \ ,
\\
&
\l_0\l_1^{-1}\del_+ A_- - \del_-B_+ + \l_0 \l_3^{-1} \del_+ B_-
= \l_0 [B_+,\l_3^{-1}B_-] + \l_0 [B_+,\l_1^{-1} A_-]\ .
\end{split}
\ee
These are written solely in terms of the gauge fields and the group elements are implicitly present
via \eqn{abp} and \eqn{abm}.

\no
The actions \eqn{gaufix} and \eqn{defactigen} as well as the set of equations of motion \eqn{eomAinitial1} and \eqn{eomAinitial2} are invariant under the $\mathbb{Z}_2$-symmetry
\be
k_1 \leftrightarrow k_2\ ,\quad  \l_1\leftrightarrow \l_2\ ,\quad \l_3 \leftrightarrow \l_4\ ,
\quad A_\pm  \leftrightarrow  B_\pm \ ,\quad g_1  \leftrightarrow  g_2\ .
\ee
Due to this symmetry we may take $k_1\leqslant k_2$ with no loss of generality. 
In addition, the actions are invariant under the parity transformation
\be
\begin{split}
&
+ \leftrightarrow -\ , \quad 
A_+  \leftrightarrow  A_- \ ,\quad B_+  \leftrightarrow  B_-\ ,
\quad g_i\to g_i^{-1}\ ,\quad i=1,2\ ,
\\
&
 \l_1\leftrightarrow \l_2^T\ ,\quad \l_3 \to \l_3^T\ , \quad \l_4 \to \l_4^T\ ,
\end{split}
\ee
upon which the equations \eqn{eomAinitial1} and \eqn{eomAinitial2} are interchanged.

We note the following interesting case.  
If we choose
\be
\begin{split}
&
\l_1^{-1}=\l_2^{-1} = {1\ov 4} (1-\l^{-1})(\l_0+\l_0^{-1})\ ,
\\
&
\l_3^{-1}= \l_4^{-1} = {1\ov 4} \l_0^{-1}(\l^{-1}-1) + {1\ov 4} \l_0 (\l^{-1}+3)\ 
\end{split}
\ee
and after redefining $A_\pm\to A_{1\pm}$ and $B_\pm \to A_{2\pm}$ then the action \eqn{gaufix} become that in Eq. (2.1) of \cite{Sfetsos:2017sep}.
It has been shown in that work that the corresponding $\s$-model action is the $\l$-deformation $G_{k_1}\times G_{k_2}/G_{k_1+k_2}$ coset CFT model and moreover it is an integrable one. Various other properties of this special $\l$-deformed models were worked out extensively in 
 \cite{Sfetsos:2017sep}.

In the rest of the paper we will restrict ourselves to the isotropic case in which all coupling 
matrices are proportional to the identity, reserving nevertheless the same symbol of the 
proportionality constant, i.e.
\be
(\l_i)_{ab} =  \l_i \d_{ab}\ ,\quad \forall \ i=1,2,3,4\ . 
 \ee 

\section{ Truncation to a two-parameter integrable model}

In  this section we discuss in detail a two-parameter model which arises by taking the limit $\l_2,\l_3\to \infty$. 
We will see below in the calculations of the $\b$-functions of the model that this is a consistent truncation of the full theory. 
From \eqn{gaufix} we see that there is a term mixing $B_+$ and $A_-$. Hence,
we expect that the resulting $\s$-model will describe self-interactions as well mutual ones. 
Indeed, the perturbation from the conformal 
point will be driven by the first and fourth bilinear in \eqn{alll}. 
Simple quantum field theoretical arguments show that the inclusion of third bilinear is not self-consistent at the quantum level and one necessarily has to include the remaining fourth bilinear as well.

\subsection{Truncation of the action and the equations of motion}

In the limit $\l_2,\l_3\to \infty$. the action \eqn{defactigen} takes the form
\be
\begin{split}
&  S_{k_1,k_2}(g_1,g_2) = S_{k_1}(g_1) + S_{k_2}(g_2)
\\
& \qq\qq + {k_1\ov \pi} \int  d^2\s\ J_{1+} (\l_0^{-1}\l_4^{-1}\mathbb{1}-D_1^T)^{-1} J_{1-}
\\
&\qq\qq - {k_2\ov \pi} \int  d^2\s\ J_{2+} D_2 J_{2-}
\\
& \qq\qq
+{k_2\ov \pi} \l_0 \l_1^{-1} \int  d^2\s\ J_{2+} D_2 (\l_0^{-1}\l_4^{-1}\mathbb{1}
-D_1^T)^{-1} J_{1-}
\ .
\end{split}
\label{dlimit}
\ee
One observes that combining $S_{k_2}(g_2)$ and the third line we obtain
the WZW model action $S_{-k_2}(g_2^{-1})$ which has negative signature.\footnote{This is expected since the conditions for having a Euclidean signature for the PCM part of the
action \eqn{oractt} are $E_1+E_4>0$ and $E_1 E_4-E_2 E_3>0$. For the two parameter model in question the limit $\l_2,\l_3\to \infty$ corresponds to setting $E_2=0$ and $E_4=-k_2$.
Then, these conditions simplify to $E_1>k_2$ and $E_1 k_2<0$ which are impossible to satisfy for $k_2>0$. Hence, even for the original action \eqn{gaufix} demanding Euclidean signature
requires flipping the sign of $k_2$. 
}
To remedy the situation we perform the following redefinition of the couplings and analytic continuation in
the specified order  
\be
\label{redeff}
\begin{split}
& \l_4= \l_0^{-1} \l \ ,\quad \l_1 = \l_0 \l \tilde \l^{-1}\ ,
\\
&{\rm then\ let}\ \ k_2\to -k_2\ ,\quad   g_2\to g_2^{-1} \ .
\end{split}
\ee
Then the action \eqn{dlimit} becomes
\be
\begin{split}
&  S_{k_1,k_2}(g_1,g_2) = S_{k_1}(g_1)
+ {k_1\ov \pi} \int  d^2\s\ J_{1+} (\l^{-1}\mathbb{1}-D_1^T)^{-1} J_{1-} + S_{k_2}(g_2)
\\
& \qq\qq\qq \
+ {k_2\ov \pi}\tilde \l \l^{-1}  \int  d^2\s\ J_{2+} (\l^{-1}\mathbb{1}-D_1^T)^{-1} J_{1-}\ .
\end{split}
\label{dlimit2}
\ee
The first line is the original $\l$-deformed model and a WZW model,
whereas the second line represents their mutual interaction.
Since the matrix $D_1$ is orthogonal it has eigenvalues lying on the unit circle. Therefore,
to avoid singularities we restrict to $-1< \l<1$.
Moreover, examining the determinant and the trace of the metric that can be extracted from \eqn{dlimit2} we find that Euclidean signature is guaranteed provided that the parameters $\l$ and $\tilde \l$ are such that they lie within 
the ellipsis, i.e.
\be
\label{bounds}
\l^2 + {k_2\ov k_1} \tilde \l^2 < 1\ , 
\ee
in addition to $k_1,k_2$ being positive integers. Note that since we have broken the $\mathbb{ Z}_2$-symmetry by turning off two of the possible interacting terms between gauge fields in 
we cannot assume with no generality loss that one of the levels is larger than the other since that would have 
been a restriction of the possible parametric space. This will be important when we discuss 
the RG flows equations and the associated fixed points in section 4.

\no
For small values of $\l$ and $\tilde \l$ we have that
\be
\begin{split}
& S_{k_1,k_2}(g_1,g_2) = S_{k_1}(g_1) + S_{k_2}(g_2)
\\
&
\qq \qq + {k_1\ov \pi}\l \int  d^2\s\ J_{1+} J_{1-} + {k_2\ov \pi}\tilde \l   \int  d^2\s\ J_{2+} J_{1-}+ {\cal O}(\l^2,\l\tilde \l) \ .
\end{split}
\label{dlimit3}
\ee
Hence we have simultaneously mutual as well as self-interactions between two WZW models.
Thus \eqn{dlimit2} is the corresponding exact effective action in which 
all loop effects in $\l$ and $\tilde \l$ are taken into account. 
By construction gravity is trusted for small curvatures which is 
warranted as long as $k_1,k_2\gg 1$. 

\no
One can show that the action \eqn{dlimit2} has the following non-perturbative in parameter space symmetry
\be
g_1\to g_1^{-1}\ ,\quad \l\to {1\ov \l}\ ,\quad k_1\to -k_1\ ,\quad \tilde \l\to \tilde {\l\ov \l}\ .
\label{acctin}
\ee
This is an extension of the similar symmetry for the original $\l$-deformed theory \cite{Sfetsos:2013wia} found in \cite{Kutasov:1989aw,Itsios:2014lca} and of the similar ones in \cite{Georgiou:2016urf,Georgiou:2017jfi} and \cite{Sfetsos:2017sep}. 
This symmetry mixes the two coupling constants and it will be a symmetry of the $\b$-functions
which we will compute below in section 4. Moreover, it is expected to be a symmetry of the
anomalous dimensions and of the correlation functions of the various operators in the theory as it happened 
in analogous computations in \cite{Georgiou:2015nka,Georgiou:2016iom} and \cite{Georgiou:2016zyo}.

\no
The equations of motion for this two-parameter $\s$-model case can be obtained by
taking the limit $\l_2,\l_3\to \infty$ in the equations
\eqn{eomAinitial1} and \eqn{eomAinitial2}. The result is given by
\be
\begin{split}
\label{eomAfinal2}
&
 \l_0 \l_1 \l_4 \del_+ A_- - \l_4 \del_- B_+ - \l_1 \del_- A_+= \l_4 [B_+,A_-] + \l_1 [A_+,A_-]\ ,
\\
&\del_+ B_- =0\ ,
\\
&
  \del_+ A_- - \l_0\l_4 \del_- A_+  =  [A_+,A_-]\ ,
\\
&
 \l_0   \del_+ A_- - \l_1\del_- B_+ = \l_0 [B_+,A_-]  \ .
\end{split}
\ee
\no
After the redefinitions \eqn{redeff} these  become
\be
\begin{split}
\label{eomAfinal1r}
&
 -{k_1\ov k_2}\l  \del_+ A_- - \tilde \l  \del_- B_+  + {k_1\ov k_2} \del_- A_+= \tilde \l [B_+,A_-] -{k_1\ov k_2}  [A_+,A_-]\ ,
\\
&\del_+ B_- =0\ ,
\\
&
  \del_+ A_- - \l \del_- A_+  =  [A_+,A_-]\ ,
\\
&
\tilde  \l   \del_+ A_- - \l\del_- B_+ = \tilde\l [B_+,A_-]  \ .
\end{split}
\ee
Hence, the field $B_-$ decouples from the rest of the equations. 

\no
For $\tilde \l=\l$ the above system can be further simplified to
\be
\begin{split}
\label{eomAfinal1rwr}
&
 \left(1+{k_2\ov k_1}\right) \l  \del_+ A_- -   \del_- A_+=    [A_+,A_-]\ ,
\\
&\del_+ B_- =0\ ,
\\
&
  \del_+ A_- - \l \del_- A_+  =  [A_+,A_-]\ ,
\\
&
 \del_+ A_- - \del_- B_+ = [B_+,A_-]  \ .
\end{split}
\ee
In that case the first and third equations are enough to determine $A_\pm$.
Then the last equation can be integrated in order to obtain $B_+$.

\no
These systems of equations, in particular \eqn{eomAfinal1r}, will be used next to show integrability of our two-parameter model.

\subsection{Integrability}

We will show that the above theory with two independent couplings $\l$ and $\tilde \l$ is integrable.
To achieve this goal we should be able to derive the four equations of motion
\eqn{eomAfinal1r} from a Lax pair containing a spectral parameter.
However, as it has been already mentioned, the second equation among them decouples since $B_-$ does not appear in any of the rest three equations. Furthermore, this equation implies 
chirality for $B_-$ hence implying an infinite number of conserved charges which 
can be constructed solely from $B_-$.
It is, thus, enough to determine a Lax connection for the remaining three equations of motion.

To proceed we will assume that the Lax connection  takes the following form
\be
\begin{split}
\label{Lax}
\cL_+=u  A_+ + v B_+,\qquad
\cL_-=\l^{-1}w A_- ,
\end{split}
\ee
where $u,v$  and $w$ are constants depending  
on the couplings $\l$ and $\tilde \l$, as well as on the WZW levels $k_1$ and $k_2$ and the
spectral parameter.
The deformation parameter $\l$ has been introduced in the above expression 
for convenience. From the Lax equation
\be\label{Lax-gen}
\partial _+\cL_--\partial _-\cL_+ -[\cL_+,\cL_-]=0\ ,
\ee
one then obtains
\be
\begin{split}\label{Lax1}
\l^{-1}w \partial _+  A_-   -u \partial _-  A_+ - v\partial _- B_+ - \l^{-1} u w[ A_+,  A_-] - \l^{-1} vw[B_+,  A_-]=0\ .
\end{split}
\ee
We may solve the system \eqn{eomAfinal1r} in terms of the derivatives of the gauge fields to obtain  that
\be
\begin{split}
\label{eom1-sol}
& \partial_+A_-  ={1\ov  \D}
\Big(k_1 (1-\l)[A_+,A_-] + k_2 \tilde \l (\l-\tilde \l)[B_+,A_-] \Big)\ ,
\\
& \partial_-A_+  = {  1 \ov  \l\D}
\Big(\big(k_2\tilde \l^2 -k_1\l (1-\l)\big)[A_+,A_-] + k_2\tilde \l (\l-\tilde \l)[B_+,A_-]\Big)\ ,
\\
& \partial_-B_+  ={ \tilde \l\ov  \l \D}
\Big(k_1 (1- \l) [A_+,A_-] + \big(k_2\l\tilde \l-k_1(1-\l^2)  \big)[B_+,A_-]\Big) \  ,
\end{split}
\ee
where we have defined the constant
\be
\label{ddd}
\D = k_1(1-\l^2) -k_2\tilde \l^2\ . 
\ee
Substituting \eqn {eom1-sol} into \eqn{Lax1} results into two algebraic equations obtained 
by equating to zero the coefficients of the commutators $[A_+,A_-]$ and $[B_+,A_-]$.
These are given by
\be\label{Lax3}
\begin{split}
& \big[k_2\tilde \l^2 -k_1(1-\l^2)\big]u w  -  \big[k_2\tilde \l^2 -k_1\l(1-\l)\big]u  
 + k_1 (1-\l)(w-\tilde \l v ) = 0 \ ,
\\
&
  \big[k_2\tilde \l^2 -k_1(1-\l^2)\big]v w - \tilde\l   \big[k_2\l\tilde \l -k_1(1-\l^2)\big]v +  
  k_2\tilde \l (\l-\tilde \l)(w-u) = 0 \  .
\end{split}
\ee
Obviously, satisfying this system leaves one parameter free among any combination of
$u,v$ and $w$ which may serve as the spectral parameter of the Lax pair in \eqn{Lax}.
An explicit solution is obtained by solving \eqref{Lax3} for $u$ and $v$ in terms of $w$ and identify the latter with the spectral parameter $z$. The result is
\be
\begin{split}
\label{Lax-final}
&\cL_+={k_1(\lambda -1)  z (z-\tilde \l) \ov Q} A_+ -{k_2 \tilde \l (\lambda -\tilde \l)(z-1) z  \ov Q}B_+,\qquad
\cL_-= \l^{-1}z  A_- \\
&Q=k_1 (\lambda -1) (\lambda  (z-1)+z) (z-\tilde \l)+k_2 \tilde \l ^2 (z-1) (z-\lambda ).  \\
\end{split}
\ee

\no
Note that for $\tilde \l= \l$ the Lax pair in \eqn{Lax-final} has vanishing coefficient for $B_+$.  
This is consistent with the fact that in \eqn{eomAfinal1rwr} only two of the equations 
involving $A_+$, $A_-$  are independent as explained in the text.

\no
This concludes the proof that the two parameter model is indeed integrable. 
As noted, a key ingredient for this proof is the fact that $B_-$ decouples form the other 
three fields which form a closed system of the equations. This is not the case for the 
more general system \eqn{eomAinitial1} and \eqn{eomAinitial2} which makes the investigation of
integrability in the four parameter model much more involved.

\subsection{The non-Abelian T-duality limit}

Near $\l=1$ we get a singularity in the manifold. However, one may zoom in by taking simultaneously the large $k_1$-limit as in \cite{Sfetsos:2013wia}. 
To do that the most convenient way we first rename $k_2$ and $g_2$ as $k$ and $g$, respectively. Then we expand for $k_1\gg 1$ as
\begin{equation}
\lambda =1 - {k \ov 2\zeta k_1}  + {\cal O}\left(1\ov k_1^2\right)\ , 
\quad g_1 = \mathbb{I} + i{k\ov 2\zeta} 
{ v_a t^a \ov k_1} + {\cal O}\left( 1\ov k_1^2 \right)  \ ,
\label{laborio}
\end{equation}
where $\zeta $ is a new coupling parameter. This leads to
\begin{equation}
\begin{split}
& J_{1\pm}^a ={k\ov 2\zeta} {\del_\pm v^{a}\ov k_1} + {\cal O}\left( 1\ov k_1^2 \right)\ ,\qquad
(D_1)_{ab} = \delta_{ab}+{k\ov 2 \zeta}\frac{f_{ab}}{k_1} 
+ {\cal O}\left( 1\ov k_1^2 \right) \ ,
\\
&{  f_{ab} =  f_{abc} v^c}\ .
\label{orrio}
\end{split}
\end{equation}
In this limit the action \eqn{dlimit2} becomes
\be
S = S_k(g)  + {k\ov  2\pi \zeta} \int d^2\s\ \del_+ v^a (\mathbb{1} +f)^{-1}_{ab}\del_- v^b  
+ {k\tilde \l \ov \pi}  \int d^2\s\
J_+^a (\mathbb{1} +f)^{-1}_{ab} \del_-v^b  \\ .
\label{nobag}
\ee
Note that Euclidean signature imposes a constraint on the parameters 
\be
\label{fjh1}
\zeta > 0 \ ,\qq \zeta \tilde \l^2 < 1\ .
\ee
This $\s$-model represents the interaction of a WZW model for a group $G$ and 
the non-Abelian T-dual of the PCM for the same group.
The original action corresponding to the interaction of the WZW model and the PCM model itself
via their respecting currents is given by
\be
S = S_k(g)  - {k\ov 2\pi \zeta}  
\int d^2\s\ \Tr(\tilde g^{-1}\del_+\tilde g \tilde g ^{-1}\del_-\tilde g)
- i {k\tilde \l\ov \pi} \int d^2\s\ J_+^a \Tr(t^a \tilde g^{-1}\del_-\tilde g)\ ,
\label{nobag2}
\ee
which also has Euclidean signature thanks to \eqn{fjh1}. 
Indeed, performing a non-Abelian T-duality transformation (following the conventions of \cite{Itsios:2013wd})
on this action  with respect to the left action on the group element  $\tilde g\in G$ we obtain \eqn{nobag}.

\no
We also mention a further consistent limit concerning  $\zeta\to 0$ involving also 
a stretching of the coordinates $v^a$. Specifically, 
\be
\label{hhh}
v^a = \sqrt{\zeta} x^a\ ,\qq \tilde \l ={\eta \ov \sqrt{\zeta}}\ ,\qq \zeta\to 0\ .
\ee
Then \eqn{nobag} becomes
\be
S = S_k(g)  + {k\ov  2\pi } \int d^2\s\ \left(\del_+ x^a \del_- x^a  
+ 2 \eta  J_+^a \del_-x^a \right)\  ,
\label{nobag3}
\ee
which represents the interaction of a WZW model action with flat space of equal dimensionality.
The model has Euclidean signature provided that $0<\eta ^2 <1$.

\no
Finally, let us note that in the non-Abelian limit \eqn{laborio} the equations of motion in \eqn{eomAfinal1r} become
\be
\begin{split}
\label{eomAfinal1rno}
&
{1\ov 2\zeta} \del_+ A_- -  \tilde \l  \del_- B_+  = \tilde \l [B_+,A_-] \ ,
\\
&\del_+ B_- =0\ ,
\\
&
  \del_+ A_- -  \del_- A_+  =  [A_+,A_-]\ ,
\\
&
\tilde  \l   \del_+ A_- - \del_- B_+ = \tilde\l [B_+,A_-]  \ .
\end{split}
\ee
Note that the limit is well defined since the seemingly infinite term arising by taking the limit  
in  the first equation \eqn{eomAfinal1r}
 has a vanishing coefficient thanks to the third equation above.

\no
It turns out that the systems \eqn{eomAfinal1rno} and  \eqn{eomAfinal1r} when 
$\tilde \l=\l$ are identical upon a certain identification. This can be seen by interchanging $A_+$ and $B_+$ and 
identifying the pairs of parameters $(2\zeta,\tilde \l) $
and $\displaystyle \Big({k_1\ov k_1+k_2},{1\ov \l}\Big)$. 
Since, non-Abelian T-duality on PCM with or without spectator fields is a canonical transformation \cite{Curtright:1994be, Lozano:1995jx,Sfetsos:1996pm} it is expected to preserve integrability, as it was shown for instance in \cite{Mohammedi:2008vd}. Hence,
we also conclude  that \eqn{nobag2} is an integrable $\s$-model as well.

\section{Renormalization group flows}

In this section we compute the $\b$-function equations for the couplings $\l$ and $\tilde \lambda$. 
In order to do so one should in principle resort to the general equations involving the RG for two-dimensional 
$\s$-models \cite{honer,Friedan:1980jf,Curtright:1984dz}. Although this task has been undertaken for the original  $\l$-deformation model of \cite{Sfetsos:2013wia} in 
\cite{Itsios:2014lca,Sfetsos:2014jfa} it is nevertheless a formidable one due to the 
enormous effort required in computing gravity tensors for the $\s$-model \eqn{defactigen} and
even for the simpler one \eqn{dlimit2}. 
However, there is an alternative method initiated in the present context in \cite{Appadu:2015nfa} 
for the isotropic case for $\l$-deformations and since it has been extended and applied to full 
generality \cite{Sagkrioti:2018rwg}. 
We will adopt this computational method and will present many details for  pedagogical reasons.

\subsection{The $\beta$-functions}

To compute the running of couplings we choose a particular configuration of the group elements $g_i$. Namely, we choose  $g_i = e^{\s^\m \th_\m^{(i)}}$, $i=1,2$, where the  matrices  $\th_\m^{(i)}$, $\m=\pm$
are constant and commuting.
Then $J_{i\pm}=-i \th_\pm^{(i)}$ and the expressions
\eqn{abp} and \eqn{abm} for the classical values for the gauge fields become\footnote{
One might object on the use of these special group elements and to what extend the result to 
which one will obtain this way will be background independent. The use is justified by the 
consistency of the end result. In addition, this method has been applied for the models of
\cite{Sfetsos:2013wia,Georgiou:2017jfi}
using arbitrary  group elements as backgrounds and at the end the result is background independent \cite{Sagkrioti:2018rwg}.} 
\be
\label{gg23}
\begin{split}
& 
A_+^{(0)}={1\ov \G}\left( (\l_0 \l_3^{-1}-1)\th_+^{(1)} -\l_0^{-1}\l_1^{-1} \th_+^{(2) }
\right)
\ ,
\\
&
B_+^{(0)}={1\ov \G}\left(-\l_0\l_2^{-1} \th_+^{(1)} +  (\l_0^{-1} \l_4^{-1}-1)\th_+^{(2)}
\right)\ ,
\\
& A_-^{(0)}={1\ov \G}\left(-(\l_0 \l_3^{-1}-1) \th_-^{(1)} +  \l_0^{-1} \l_2^{-1}\th_-^{(2)}
\right)\ ,
\\
&
B_-^{(0)}={1\ov \G}\left( \l_0 \l_1^{-1}\th_-^{(1)}-(\l_0^{-1} \l_4^{-1}-1) \th_-^{(2)}
\right)\ ,
\end{split}
\ee
where we have defined the constant $\G=(\l_0\l_3^{-1}-1)(\l_0^{-1}\l_4^{-1}-1) -\l_1^{-1}\l_2^{-1}$. For the gauge fields $A_\pm$ and $B_\pm$ the superscript denotes the fact that these are classical values for
the gauge fields.
Then the Lagrangian density corresponding to the action \eqn{defactigen} reads
\be
\cL^{(0)}={1\ov \pi} \left(\!
                        \begin{array}{cc}
                          \th_+^{(1)} & \th_+^{(2)} \\
                        \end{array}
                    \!\right) \left(\!
                                \begin{array}{cc}
                                  -{k_1\ov 2} -{k_1\ov \D}(\l_0\l_3^{-1}-1) & {k_2\ov \D} \l_0 \l_2^{-1}\\
                                   {k_1\ov \D} \l_0^{-1} \l_1^{-1} & -{k_2\ov 2} -{k_2\ov \D}(\l_0^{-1}\l_4^{-1}-1) \\
                                \end{array}
                             \! \right)
                                \left(\!
                        \begin{array}{c}
                          \th_-^{(1)} \\  \th_-^{(2)} \\
                        \end{array}
                     \! \right) \ .
\label{gg233}
\ee

We will be particularly interested in two parameter action \eqn{dlimit2}. 
The case with four couplings can be similarly worked out, but the resulting expressions are quite complicated and not very enlightening. 
Proceeding for the two parameter case we have that the classical solution to the gauge fields 
is given by
\be
\label{clasgg}
\begin{split}
& A_+^{(0)}={\l\ov 1-\l} \left(\l \th_+^{(1)} + {k_2\ov k_1}\tilde \l  \th_+^{(2) } \right) \ ,
\\
&
B_+^{(0)}=\th_+^{(2)}\ ,
\\
& A_-^{(0)}=-{\l\ov 1-\l } \th_-^{(1)} \ ,
\\
&
B_-^{(0)}=-{1\ov 1-\l}\left( \tilde \l  \th_-^{(1)} + (1-\l) \th_-^{(2)}\right)\ .
\end{split}
\ee
The Lagrangian density corresponding to the action  \eqn{dlimit2} reads
\be
\label{clasgg1}
\cL^{(0)}=-{1\ov \pi} \left(
                        \begin{array}{cc}
                          \th_+^{(1)} & \th_+^{(2)} \\
                        \end{array}
                      \right) \left(
                                \begin{array}{cc}
                                  {k_1\ov 2} {1+\l\ov 1-\l} & 0\\
                                   k_2 {\tilde \l \ov 1-\l} & {k_2\ov 2}  \\
                                \end{array}
                              \right)
                                \left(
                        \begin{array}{c}
                          \th_-^{(1)} \\  \th_-^{(2)} \\
                        \end{array}
                      \right)\ .
\ee
We note that in obtaining \eqn{clasgg} and \eqn{clasgg1} from \eqn{gg23} and \eqn{gg233}
we have used the
redefinition \eqn{redeff} and subsequently we let $k_2\to -k_2$ and $\th_\pm^{(2)}\to
-\th_\pm^{(2)}$ (corresponding to inverting the group element $g_2$ as in \eqn{redeff}).

\no
The next step is to consider the fluctuations of the gauge fields around \eqn{clasgg} and let
\be
\begin{split}
&
A_\pm = A^{(0)}_\pm + \d A_\pm \ ,\qq B_\pm = B^{(0)}_\pm + \d B_\pm \ ,
\\
&
(\tilde A_\pm^{(0)})_{ab} =i f_{abc} (A_\pm^{(0)})_c\ ,\qq (\tilde B_\pm^{(0)})_{ab} =i f_{abc} (B_\pm^{(0)})_c \ .
\end{split}
\ee

\no
The linearized fluctuations for the classical equations of motion
 \eqn{eomAfinal1r} can be cast in the form
\be
\hat D \left(
         \begin{array}{c}
           \d A_- \\
           \d B_+ \\
           \d B_- \\
           \d A_+ \\
         \end{array}
       \right) = 0\ ,
\ee
where the operator $\hat D$ is first order in worldsheet derivatives. 
We will present the form of this operator in the Euclidean
regime and in momentum space. That means that one should analytically continue $\tau \to -i x_1 $ and rename $\s$ as $x_2$. Denoting $z=x_1+i x_2$ we have that
\be
\s^+ \to -i z\ ,\quad \s^-\to -i \bar z\ ,\quad \del_+\to i \del \ ,\quad \del_-\to i \bar \del\ .
\ee
In addition, in passing to momentum space we have, for the plane wave basis we use, that
\be
e^{i (p_+ \s^+ +  p_- \s^-)}=  e^{{i\ov 2} (p^- \s^+ + p^+ \s^-)}\to e^{-{i\ov 2}(\bar p z+ p\bar z)}\ .
\ee
Hence, the derivatives acting on the plane waves give in the Euclidean regime the following result
\be
(\del_+,\del_-) e^{i (p_+ \s^+ +  p_- \s^-)}\to \ha(\bar p,p) e^{-{i\ov 2}(\bar p z+ p\bar z)}
\ .
\ee
Taking these into account and denoting for notational convenience $\ha(\bar p,p)$ by $(p_+,p_-)$, we have that $ \hat D = \hat C + \hat F $, where
\be
\hat C = \left(
           \begin{array}{cccc}
             -{k_1\ov k_2} \l  p_+ & -\tilde \l p_- & 0 & {k_1\ov k_2}  p_- \\
             0 & 0 & p_+ & 0 \\
              p_+ & 0 & 0& -\l p_- \\
             \tilde \l  p_+ & -\l p_- & 0& 0 \\
           \end{array}
         \right)
\ee
and
\be
\hat F = \left(
           \begin{array}{cccc}
 \tilde\l \tilde B^{(0)}_+\!  -  {k_1\ov k_2} \tilde A_+^{(0)}   &- \tilde \l \tilde A^{(0)}_- & 0 & {k_1\ov k_2} \tilde A^{(0)}_- \\
0 & 0 & 0 & 0 \\
        \tilde A^{(0)}_+ & 0 & 0  & -  \tilde A^{(0)}_- \\
             \tilde \l \tilde B^{(0)}_+ & -\tilde \l  \tilde A^{(0)}_-  & 0 & 0 \\
           \end{array}
         \right)\ .
\ee
The effective Lagrangian of our model is then given by
\be
-\cL_{\rm eff} = \cL^{(0)} + \int^\m {d^2 p\ov (2\pi)^2} \ln (\det \hat D)^{-1/2}\ .
\ee
We are interested in the logarithmic divergence of this integral with respect to the UV mass scale 
$\m$. Therefore, we will perform a large momentum expansion of the integrand.
We need to keep only terms proportional to $\displaystyle {1\ov p_+ p_-}$
since these are the ones which, upon integration over the momenta, will give rise to a logarithmic $\ln \m$, divergence. Using the fact that
\be
\begin{split}
& \ln (\det \hat D) = \ln \det \hat C + \Tr \ln(\mathbb{1} + \hat C^{-1} \hat F)
\\
& \qq\qq  = \ln \det \hat C + \Tr (C^{-1} \hat F) -\ha \Tr(C^{-1} \hat F)^2 + \cdots  \ .
\end{split}
\ee
the only term in the above equation that will contribute as described above is the last one written.
Indeed, by isolating the momentum dependence, one can write the inverse of the matrix $\hat C$ as
\be
\hat C^{-1}= {\hat c_+\ov p_+} + {\hat c_-\ov p_-}\ .
\ee
Hence, the relevant part in $\cL_{\rm eff}$ is
\be\label{Leff}
\begin{split}
&
-\cL_{\rm eff} = \cL^{(0)} + {1\ov 8 \pi^2} \underbrace{\int^\m {d^2p \ov p_+p_-}}_{4\times 2 \pi \int^\m{dp\ov p}}  \Tr(\hat c_+ \hat F \hat c_- \hat F)
\\
& \qq \quad =  \cL^{(0)} + {1\ov 2\pi } \ln\m^2 \ \Tr(\hat c_+ \hat F \hat c_- \hat F) \ ,
\end{split}
\ee
where we have finally substituted the complex momenta,
i.e. $p_+=\bar p/2$ and $p_-=p/2$ and we have rewritten $d^2p =2\pi p dp$ in
polar coordinates and for angle independent integrands.
Next we demand that this action is $\m$-independent, i.e. $\del_{\ln \m^2} \cL_{\rm eff}=0$.
To leading order in $k_1$ and $k_2$ this derivative acts only on the coupling constants
in $\cL^{(0)}$.

\no
The above formalism is quite general. Specializing to our case we get
\be
\begin{split}
&
\hat c_+  =    { 1 \ov \D}
\left(       \begin{array}{cccc}
            k_2 \l& 0 &   k_1 &
            - k_2 \tilde \l  \\
             0 & 0 &0 & 0 \\
              0 & 1 & 0& 0 \\
             0 & 0 & 0& 0 \\
           \end{array}
         \right)\ ,
\\
&
\hat c_- =  { 1 \ov \l\D}
\left(       \begin{array}{cccc}
            0 & 0 &  0 &   0  \\
            k_2\l \tilde \l & 0 & k_1\tilde \l & -k_1(1-\l^2) \\
              0 & 0 & 0& 0 \\
             k_2 \l & 0 & k_1\l^2 + k_2\tilde \l^2& -k_2\tilde \l \\
           \end{array}
         \right)\ ,
\end{split}
\ee
where $\D$ is the same constant defined in \eqn{ddd}.
Evaluating the trace in \eqn{Leff} we obtain
\be
\label{hd3k12}
\begin{split}
&
\Tr(\hat c_+ \hat F \hat c_- \hat F) ={c_G\ov (1-\l)\D^2 }
\Big[ k_1\l \left(-k_1\l(1-\l)+k_2\tilde \l^2 (1+\l-\tilde \l)\right)\th_+^{(1)} \th_-^{(1)}
\\
&\qq -k_2\tilde \l^2  \left(1+\l^2 +\l(1-\tilde \l)-\tilde \l\right)\left(k_1(1-\l)-k_2\tilde \l)\right)
\th_+^{(2)} \th_-^{(1)} \Big]\ ,
\end{split}
\ee
were we have use that $\Tr(\tilde A_+^{(0)} \tilde A_-^{(0)})= c_G (A_+^{(0)})^a
(A_-^{(0)})^a$ and where $c_G$ is the eigenvalue of the quadratic Casimir in the adjoint 
representation defined as $f_{acd}f_{bcd}=c_G \d_{ab}$.
Subsequently we have substituted the classical expressions \eqn{clasgg}.
On the other hand
\be
{d\ov d\ln \m^2} \cL^{(0)} = -{k_1\ov \pi} {\b_\l\ov (1-\l)^2 }\th_+^{(1)}\th_-^{(1)}
-{k_2\ov \pi} \left({\tilde \l \b_\l\ov (1-\l)^2} + {\b_{\tilde \l}\ov 1-\l}\right)
\th_+^{(2)}\th_-^{(1)}\ ,
\ee
which has precisely the same structure as \eqn{hd3k12}. This observation is closely related to the fact that 
truncating  the full theory to the one with two couplings is consistent with the RG equations.
Imposing the condition $\del_{\ln \m^2} \cL_{\rm eff}=0$, we 
get that the $\b$-functions are given by
\be
\label{systrg1}
\b_\l(\l,\tilde \l) = -c_G{ \l(1-\l) }{ k_1 \l (1-\l) -k_2 \tilde \l^2 (1+\l-\tilde \l)\ov 2  (k_1(1-\l^2)-k_2\tilde \l^2)^2}
\ee
and
\be
\label{systrg2}
\b_{\tilde \l}(\l,\tilde \l) =-c_G\tilde \l (1-\tilde\l)
{k_1(1-\l)\left(\tilde \l-\l(\l-\tilde\l)\right)-k_2\tilde \l^2  \ov 2  (k_1(1-\l^2)-k_2\tilde \l^2)^2}\ ,
\ee
each of which depends on both couplings as expected and argued for below \eqn{alll}.

We mention in passing  that the above expressions are obtainable from the most general RG flow 
equations for non-isotropic single $\l$-deformations  \cite{Sagkrioti:2018rwg}. 
This can be achieved by 
embedding the currents $J_{1\pm}$ and $J_{2\pm}$ into a single current ${\cal J}_\pm
=(J_{1\pm},J_{2\pm})$ and subsequently setting $k_1=k_2=1$. It turns out that, after some 
appropriate rescalings of the currents, the dependence on the levels is reinstated by letting the structure constants to be $(f_{abc}/\sqrt{k_1},f_{abc}/\sqrt{k_2})$. In the two coupling model case and in the
above basis, the deformation matrix reads 
\be
\displaystyle \L=
\left(       \begin{array}{cc}
            \l \mathbb{1}& 0   \\
             \l_0^{-1}\tilde \l \mathbb{1} &  0\\
           \end{array}
         \right)\ .
    \ee
We have checked using eq. (2.11) of \cite{Sagkrioti:2018rwg}
that \eqn{systrg1} and  \eqn{systrg2} are indeed reproduced. Because in the derivation of this
general equation the inverse of  the deformation matrix is used and $\L$ above is non-invertible, we preferred to perform the independent analysis presented in this subsection.

\subsection{Properties of the RG flow}

The above $\b$-function equations are invariant under the non-perturbative symmetry
\be
\label{duallii}
k_1\to -k_1\ ,\qq \l\to {1\ov\l}\ ,\qq \tilde \l\to {\tilde \l\ov \l}\ ,
\ee
as expected from the corresponding invariance \eqn{acctin} of the action. The transformation 
involves a mixing of the two parameters consistent with the fact that the system consisting of \eqn{systrg1} and \eqn{systrg2} is coupled.

\no
We have the following interesting limiting cases which also may serve as a check of our results:

\no
$\bullet $ It is consistent with the RG-flow equations to set $\tilde \l=0$. In this limit
\be
\tilde \l = 0:\qquad \b_\l = -{c_G\ov 2k_1} {\l^2 \ov  (1+\l)^2}\ ,
\ee
which is the $\b$-function for the original $\l$-deformed model 
found in \cite{Kutasov:1989dt,Itsios:2014lca}. This is consistent with the fact that the action \eqn{dlimit2}
becomes the sum of the $\l$-deformed action with level $k_1$ and that for the WZW model
$S_{k_2}(g_2)$.

\no
$\bullet $
Next consider setting $\l=0$ which is also a mathematically consistent truncation. 
Then if we redefine $\zeta=\l_0^{-1}\tilde \l$, where as usual
$\l_0=\sqrt{k_1\ov k_2}$, we have that
\be
\label{ggh1}
\l = 0:  \qquad \b_{\zeta } = -{c_G\ov 2 \sqrt{k_1k_2}}
 {\zeta^2 (\zeta-\l_0)(\zeta-\l_0^{-1})\ov  (1-\zeta^2)^2}\ ,
\ee
as it should be since the action \eqn{dlimit2} in that limit becomes the action found in
\cite{Georgiou:2016urf,Georgiou:2017jfi}
for two mutually interacting WZW models with only one possible coupling turned on. 

\no
$\bullet $
We may also consistently truncate the system by letting $\tilde \l=\l$. Then 
\eqn{systrg1} and \eqn{systrg2} degenerate to one equation given by
\be
\tilde \l=\l: \qq \b_\l = -c_G \l^2 (1-\l) {k_1 -(k_1+k_2) \l\ov 2 (k_1 -(k_1+k_2)\l^2)^2}\ .
\label{kjhk1}
\ee
This expression can also be obtained in the following alternative way. When the two couplings are equal,
it can be seen from \eqn{dlimit3} that the perturbation is of the form
\be
\begin{split}
&
{\l\ov \pi}(k_1 J_{1+} + k_2 J_{2+})J_{1-} = {{\xi}\ov \pi}\sqrt{(k_1+k_2)k_1}\
\J_+ J_{1-}\ ,
\\
& \xi = \sqrt{k_1+k_2\ov k_1}\l\ ,\qq  J_+= {1\ov k_1+k_2}(k_1J_{1+}+k_2 J_{2+})\ .
 \end{split}
 \ee
 In our normalization $k J_\pm$ generate current algebras at level $k$.
 Therefore,  $(k_1+k_2)J_+$ is a current algebra at level $k_1+k_2$.
 Hence we may use  \eqn{ggh1} for
 $\l_0= \sqrt{k_1\ov k_1+k_2}$ to obtain the RG equation for the coupling $\xi$.  
 The result is indeed given by \eqn{kjhk1} after the appropriate rescaling is taken into 
 account.

\no
$\bullet $ For equal levels we have that
\be
\begin{split}
& k_2=k_1: \qq \b_\l(\l,\tilde \l) = -c_G{ \l(1-\l) }{ \l (1-\l) - \tilde \l^2 (1+\l-\tilde \l)\ov 2  k_1(1-\l^2-\tilde \l^2)^2}\\
&\qq \qq \qq\b_{\tilde \l}(\l,\tilde \l)=\b_{\l}(\tilde\l, \l),
\end{split}
\ee
that is the expression for $\b_{\tilde \l}$ is obtained by interchanging $\l$ and $\tilde \l$. The above $\b$-functions are in agreement with eq. (3.2) of
\cite{LeClair:2001yp} (after identifying $g_1=4 \l$ and $g_2=4\tilde \l$). In this work the result 
was found by ressuming the perturbation series for the linearised action \eqn{dlimit3}.

\no
$\bullet $ In the non-Abelian limit \eqn{laborio} the RG-flow equations \eqn{systrg1} and
\eqn{systrg2} become
\be
\b_\zeta = -{c_G\ov 4k} \zeta^2 { 1-2\zeta \tilde \l^2 (1-\tilde \l)\ov  (1-\zeta \tilde \l^2)^2}
\ee
and
\be
\b_{\tilde \l} =-{c_G\ov 4k} \tilde \l (1-\tilde\l) \zeta
{2 \tilde \l-1-2\zeta\tilde \l^2  \ov (1-\zeta \tilde \l^2)^2}\ .
\ee
Finally, in the further limit \eqn{hhh} the $\b$-function for $\zeta$ is automatically satisfied,
whereas that for the coupling constant $\eta$ becomes
\be
\beta_\eta = {c_G \ov 2 k} {\eta^3 \ov (1-\eta^2)^2}\ . 
 \ee 
 This corresponds to the $k_1\to \infty$ (or $k_2\to \infty$) limit of \eqn{ggh1} as one 
 expects since the corresponding actions become identical.

\subsection{RG fixed points}
In this section, we elaborate on the structure of the RG equations by identifying the RG fixed points and by presenting several figures exhibiting the flow of the theory in the  $(\l,\tilde \l)$ plane.

For generic values for $k_1$ and $k_2$ there are six points at which the $\b$-functions \eqn{systrg1} and \eqn{systrg2}
vanish simultaneously. Specifically, these are located at the points $(\l^*,\tilde\l^*)$ given by
\be
\begin{split}
&
F_1: (0,0)\ ,\qq F_2: \Big(0,{k_1\ov k_2}\Big)\ ,\qq F_3: (0,1)\ ,\qq 
F_4: \Big({k_1\ov k_1+k_2},{k_1\ov k_1+k_2}\Big)\ ,
\\
&F_5: \Big(1-{k_2\ov k_1},1\Big)\ ,\qq F_6: (1,1)\ .
\end{split}  
\ee
Note that points $F_4$ and $F_5$ are related by the transformation \eqn{duallii} whereas
$F_6$ is left invariant.  When  $k_1=k_2$, then $F_2$  and $F_3$ and $F_5$ degenerate to the same point. 
Note that there is a seemingly seventh zeroth of the $\b$-functions at the point $(1,0)$. However, this limit in the 
$\b$-functions is not well defined since one gets different results depending on the order one takes the limit.

\no
The RG flows and the above fixed points are depicted at Figs. 1,2 and 3 for the cases where $k_1< k_2$,  $k_1>k_2$ and $k_1=k_2$, respectively. 
In each figure the left part encodes RG flows in the entire $(\l,\tilde\l)$-plane and the ellipsis (or 
circle) \eqn{bounds}
denotes the border within which the signature of the $\s$-model \eqn{dlimit2}
remains Euclidean. The right part is just a zooming of the first quadrant.

Below we describe in detail the RG flows and the corresponding fixed points.
Note that the point $F_6$ always lies outside the ellipsis bounding the region of Euclidean signature regime no matter what the ratio of $\displaystyle k_1\ov k_2$ is, whereas the points 
$F_1$ and $F_4$ always lie in. 
In addition, to understand these RG flows we have computed the stability matrix defined
as $H_{ij} = \del_{\l_j}\b_{\l_i}$ at the relevant fixed points.
For each fixed point within the Euclidean domain regime the eigenvalues of the stability matrix $(H_1,H_2)$  are given by the values in the parenthesis below
\be
\begin{split}
& F_1: (0,0)\ ,\qq F_2: \Big({1/2\ov  k_2-k_1},{1/2\ov  k_2-k_1 }\Big)\ ,
\qq F_3: \Big({1/2\ov  k_1-k_2},0\Big)\ ,
\\
& 
F_4: \Big({1\ov 2  k_2},-{1/2\ov  k_1+k_2}\Big)\ ,
\qq F_5: \Big({1\ov 2  k_2},{1/2\ov  k_1-k_2 }\Big)\ .
\end{split}
\ee
We recall that two positive (negative) eigenvalues corresponds to an IR stable (unstable) fixed point. The corresponding directions are then irrelevant and relevant, respectively. 

\no
Specifically, we have that:

\no
\underline{ $ {\rm Case} \,\,\,k_1<k_2$ :} 
This case is depicted at Fig.1.
The three zeros of the $\b$-functions inside the ellipsis bounding the region of Euclidean signature regime are:

\no
$\bullet$
The point  $F_1$ which is the CFT point corresponding to the CFT $G_{k_1}\times
G_{k_2}$ as it is clear from \eqn{dlimit3}. 

\no
$\bullet$ The point $F_2$ at which the action \eqn{dlimit2} becomes a sum of two WZW actions, i.e.
$S_{k_1}(g_1 g_2)+S_{k_2-k_1}(g_2)$,  
corresponding to the CFT $G_{k_1}\times G_{k_2-k_1}$ as in \cite{Georgiou:2017jfi}.
Clearly this is an IR stable point. 

\no
$\bullet$ The point $F_4$, which has one relevant and one irrelevant direction. It is not clear to what CFT this point correspond to.

\no
\underline{ $ {\rm Case} \,\,\,k_1>k_2$ :} 
This case is depicted at Fig.2.
The four zeros of the $\b$-functions inside the ellipsis bounding the region of Euclidean signature regime are:

\no
$\bullet$ The point  $F_1$ which is the CFT point corresponding to the CFT $G_{k_1}\times
G_{k_2}$ as in the previous case.

\no
$\bullet$ The point $F_3$ at which the action \eqn{dlimit2} becomes a sum of two WZW actions and the corresponding CFT is $G_{k_2}\times G_{k_1-k_2}$. It has one irrelevant direction.

\no
$\bullet$ The point $F_4$, with one relevant and one irrelevant direction. It is not clear to what CFT this point correspond to.

\no
$\bullet$ The point with $F_5$ which clearly is an IR stable point.
It is also not clear to what CFT this point correspond to.

\no
\underline{ $ {\rm Case} \,\,\,k_1=k_2$ :} 
This case is depicted at Fig.3.
The two zeros of the $\b$-functions inside the ellipsis bounding the region of Euclidean signature regime are:

\no
$\bullet$ The CFT point  $F_1$ as in the two previous cases.

\no
$\bullet$ The point $F_4$ similar to the two previous cases.

\no
Note that, the point $F_2$ (the same now as $F_3$ and $F_5$) is a singular one 
since the action  \eqn{dlimit2} becomes $S_{k_1}(g_1 g_2)$ which is of dimensionality $\dim G$ instead of 2 $\dim G$.

\no
\underline{ Non-Abelian limit :} 
The RG flow is depicted in Fig. 4. In the neighborhood of the line $\l=1$  the 
$\s$-model becomes strongly coupled and it makes sense to perform the zoom in 
non-Abelian limit \eqn{laborio}.  The physical 
Euclidean signature region is bounded, according to \eqn{fjh1},
between the two red lines and the $\tilde \l$-axis. 

 \begin{figure}[h]
\label{betaplot1}
\begin{center}
\hskip -.4 cm
\includegraphics[height= 8 cm,angle=0]{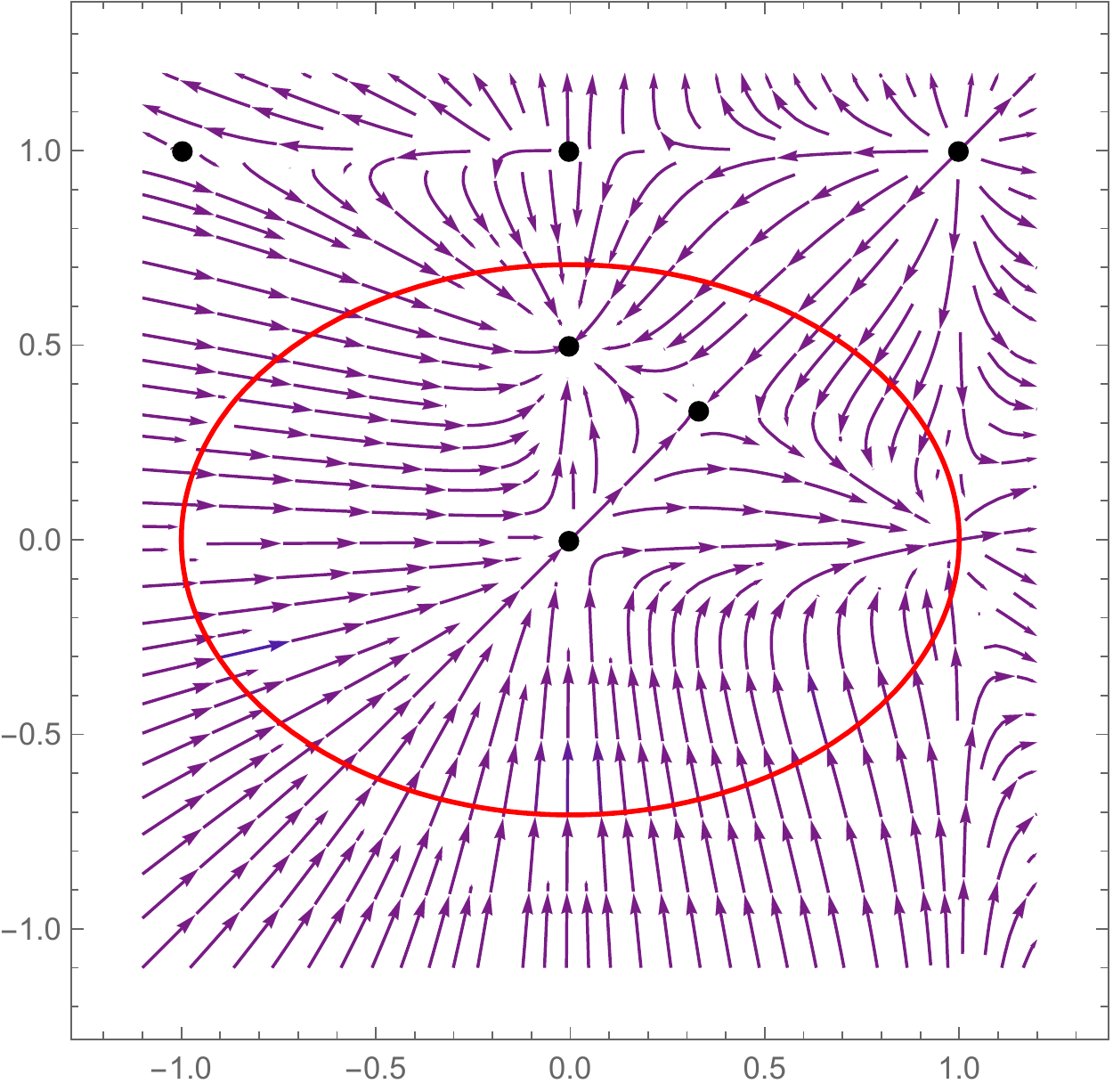}
\hskip -.16 cm
 \includegraphics[height= 8 cm,angle=0]{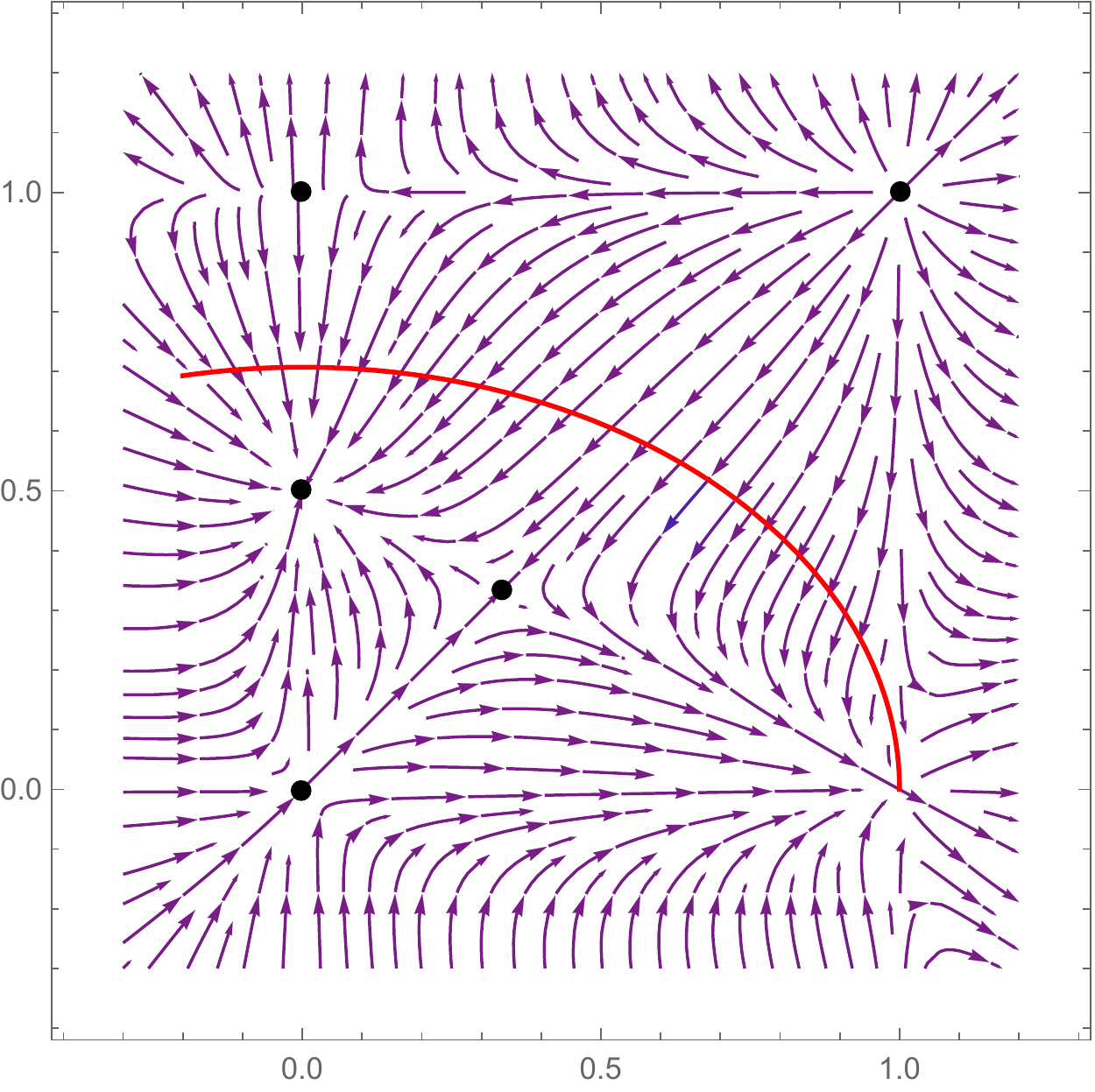}
\end{center}
\vskip -0.7 cm
\caption{RG flows in the $(\l,\tilde \l)$-plane ($\l$-horizontal) for $k_2=2k_1$: The entire
plane (left), the first quadrant (right). The region bounded by the ellipsis is the allowed region in which the signature of the metric remains Euclidean. The arrows point towards the IR.}
\end{figure}

\begin{figure}[h!]
\label{betaplot2}
\begin{center}
\hskip -.4 cm
\includegraphics[height= 8 cm,angle=0]{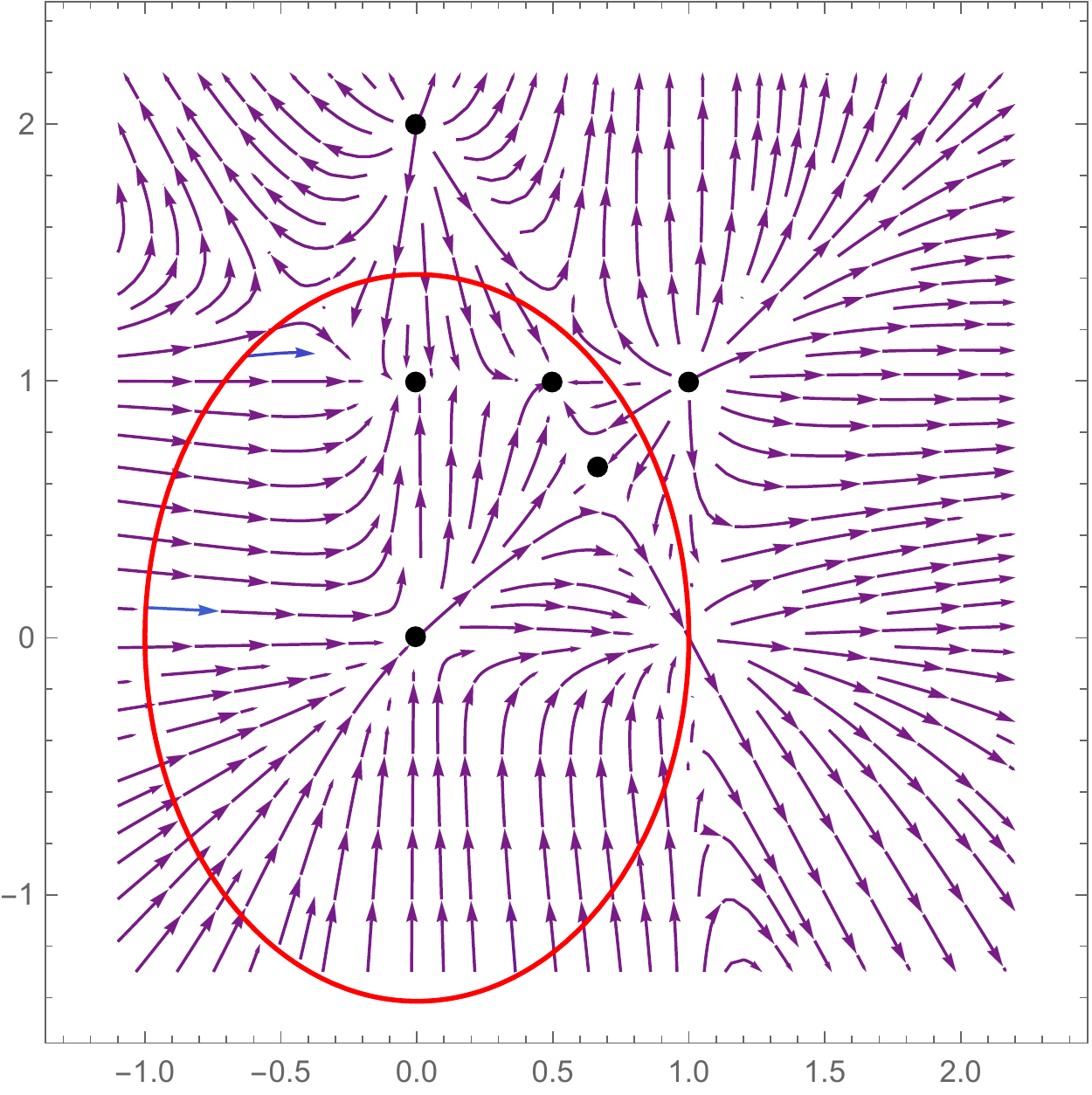}
\hskip -.16 cm
 \includegraphics[height= 8 cm,angle=0]{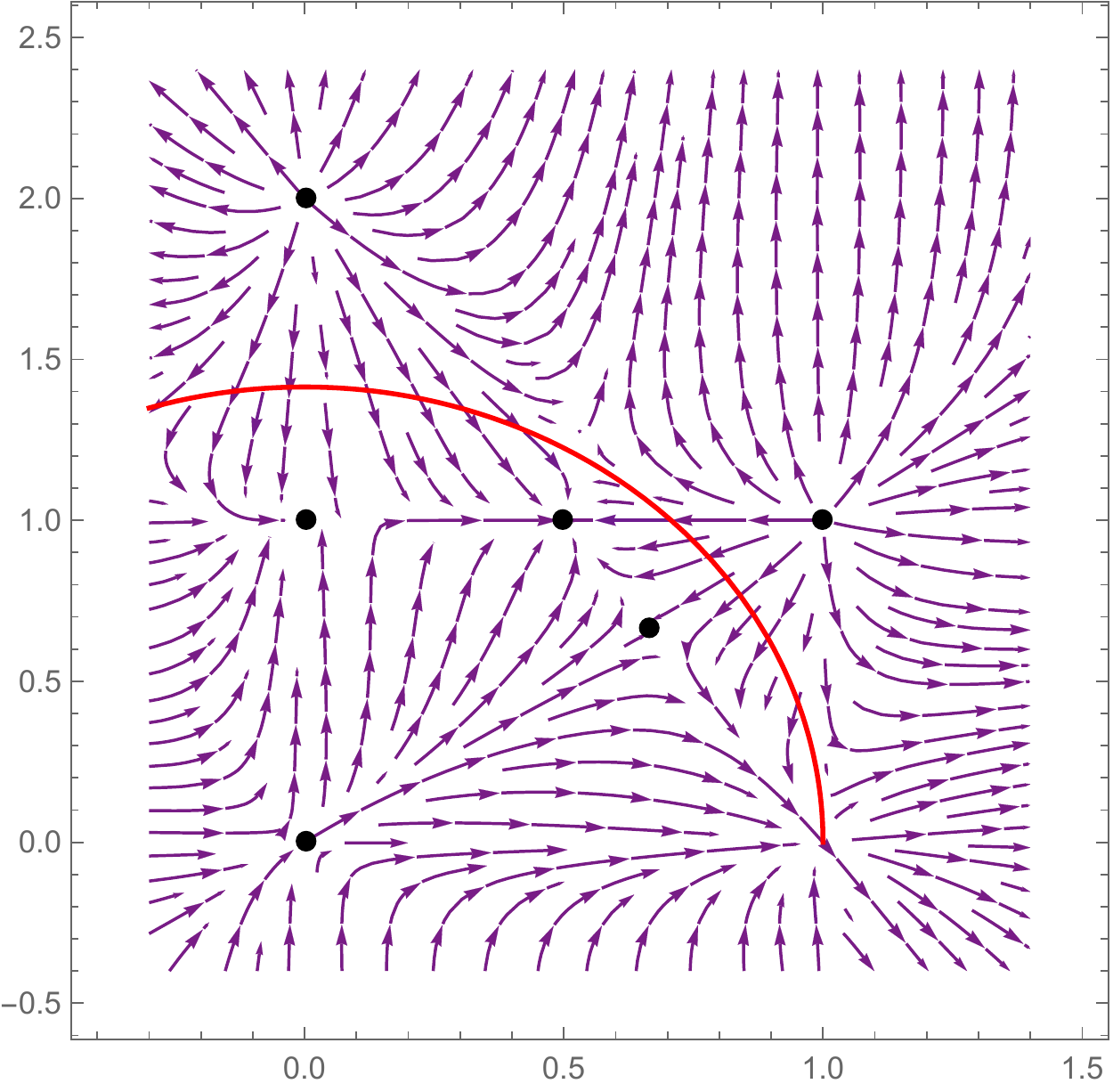}
\end{center}
\vskip -0.7 cm
\caption{RG flows in the $(\l,\tilde \l)$-plane ($\l$-horizontal) for $k_1=2k_2$: The entire
plane (left), the first quadrant (right).}
\end{figure}

\begin{figure}[h!]
\label{betaplot3}
\begin{center}
\hskip -.4 cm
\includegraphics[height= 8 cm,angle=0]{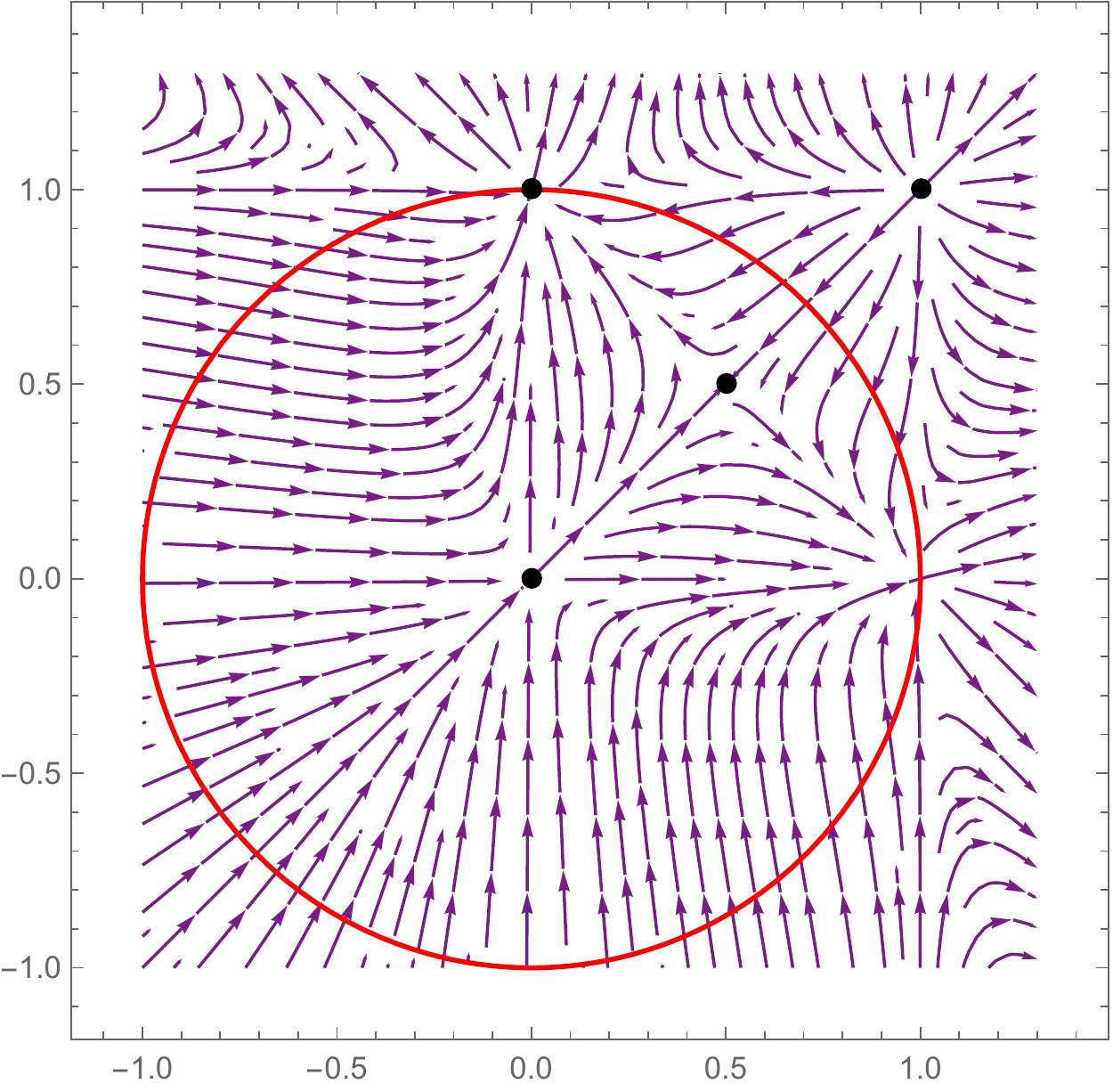}
\hskip -.16 cm
 \includegraphics[height= 8 cm,angle=0]{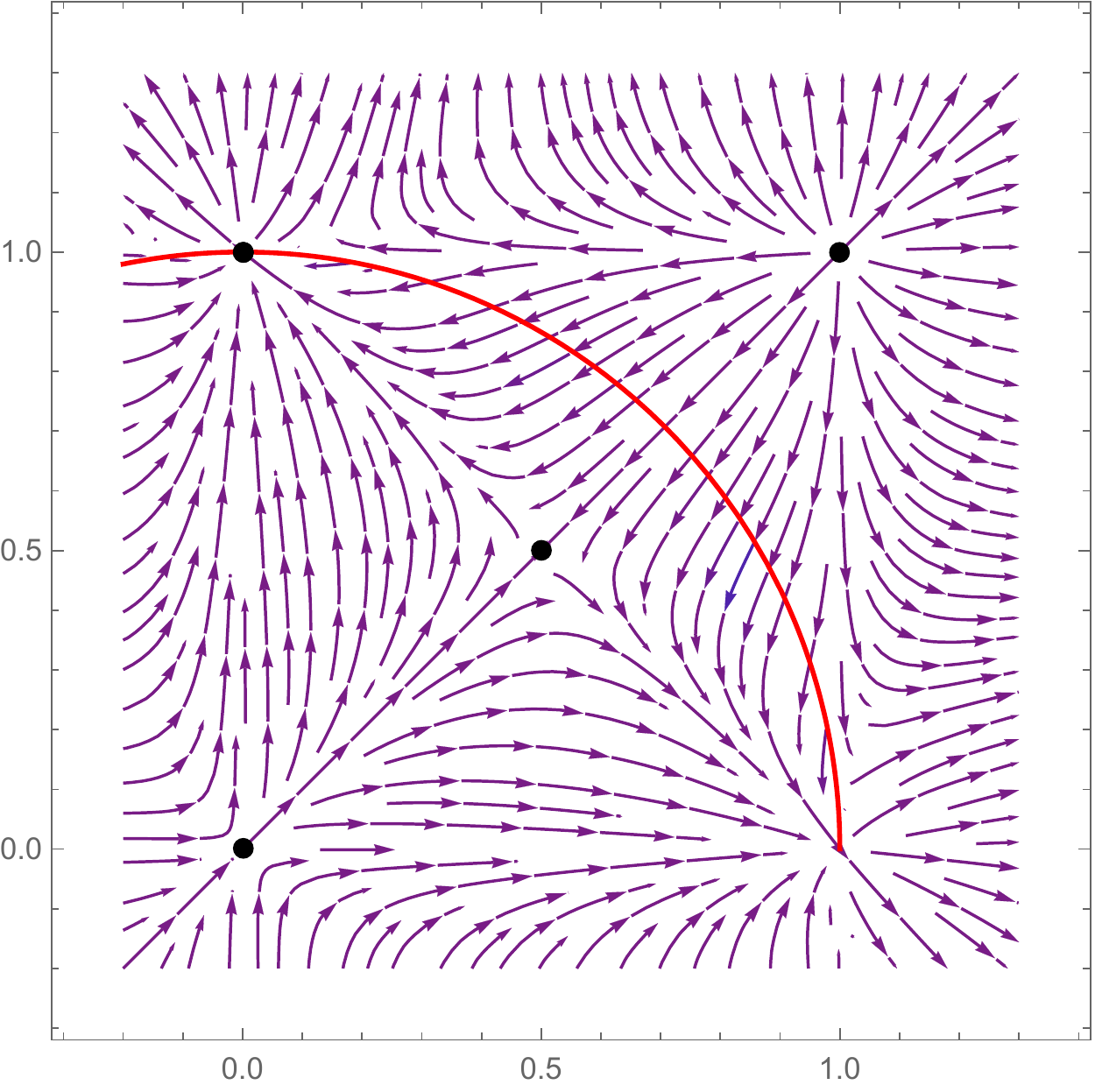}
\end{center}
\vskip -0.7 cm
\caption{RG flows in the $(\l,\tilde \l)$-plane ($\l$-horizontal) for $k_1=k_2$: The entire
plane (left), the first quadrant (right).}
\end{figure}

\begin{figure}
\label{betaplot4}
\begin{center}
\vskip -2 cm
\includegraphics[height= 8 cm,angle=0]{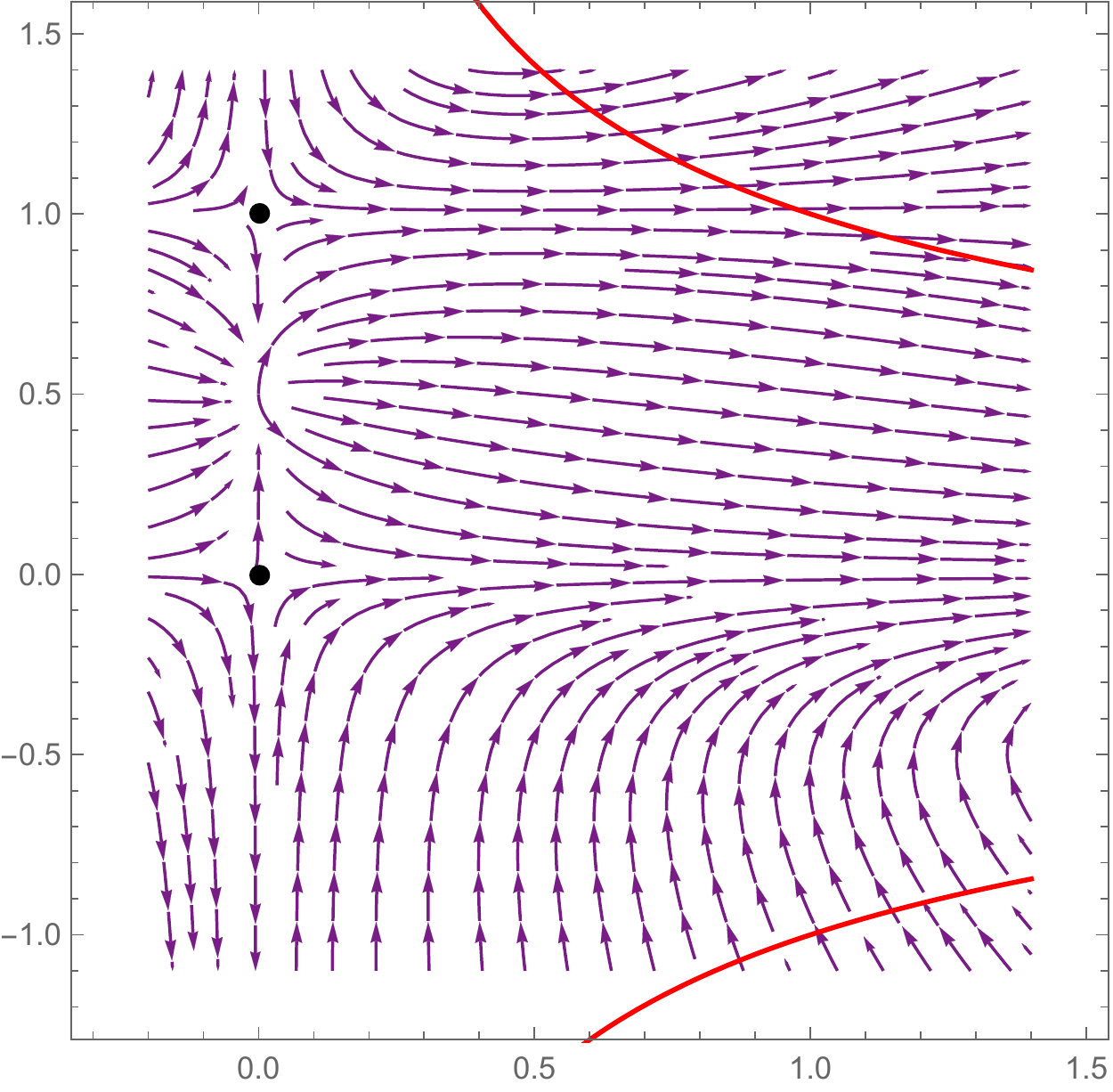}
\end{center}
\vskip -0.7 cm
\caption{RG flows in the $(\zeta,\tilde \l)$-plane ($\zeta$-horizontal) for  the non-Abelian T-dual limit.}
\end{figure}

\section{Discussion and future directions}

It is a rare and remarkable occasion when one is able to derive exact results in a Quantum Field Theory (QFT) in any number of dimensions.
In certain cases, this may be achieved in conjunction with some hidden symmetry, usually non-perturbative in nature, which the theory possesses. 
Such an example is the maximally supersymmetric gauge theory in 4-dimensions, i.e. ${\cal N}=4$ SYM, where integrability is the symmetry playing 
the instrumental role. 

Recently, an effective and rather effortless method was developed in order to obtain exact results in certain two-dimensional QFTs. The theories under consideration are conformal field theories 
of the WZW type perturbed by current bilinear operators. The method relies on the construction of the corresponding all-loop effective actions
for these theories \cite{Sfetsos:2013wia,Georgiou:2016urf,Georgiou:2017jfi}.  One then uses these effective actions to determine certain non-perturbative  symmetries in the space of couplings.
Making the plausible assumption that the symmetries of the action are inherited by the observables of the theory, one uses low-order perturbation theory, as well as the non-perturbative  symmetries in order to derive 
exact expressions for the observables. This programme was initiated and implemented in a series of papers  mentioned in detail in the introduction in which exact expressions were obtained for the $\b$-functions, for the anomalous dimensions of current and 
primary operators, as well as for the 3-point correlators involving currents and/or primary operators. We note in passing that many of the examples considered are also integrable, although this property has not been essentially in the aforementioned calculations.

In this work we continue this line of research by considering a general class of models whose UV Lagrangian is the sum of two WZW models at different levels. The perturbations driving the theory off conformality  consist of current bilinears involving currents belonging to both the same and different CFTs. The all-loop effective action of these models and the corresponding 
equations of motion were constructed in section 2. 
In general these models depend on four general coupling matrices. In section 3 and for simplicity, we consider a consistent truncation of the theory in which only two of the couplings, $\l$ and $\tilde \l$, are present. Firstly, we identify a non-perturbative symmetry in the space of couplings $\l$ and $\tilde \l$. Subsequently, we proved that the theory is classically integrable by finding the appropriate Lax connection. In the same section, we also consider non-Abelian T-duality type limits for the case of the models with two couplings.
We then proceeded in section 4 to derive the exact in the couplings $\b$-functions of our model.  To this end 
we evaluate the determinant of the matrix driving the fluctuations around a classical configuration that solves the equations of motion. The expressions for the $\b$-functions of our model enjoy the aforementioned non-perturbative symmetry. The RG flow equations have a rich structure which also depends on the relative value of the WZW levels $k_1$ and $k_2$. Subsequently, we determine the fixed points of the flow and  the nature of the corresponding CFTs in most of the cases. For the cases $k_1>k_2$ and $k_2>k_1$ there is always a fixed point, in the allowed range of the space of couplings, that is an IR attractor. This is not the case for $k_1=k_2$ where all fixed points have both relevant and irrelevant directions. Lets us also mention that our models provide {\it concrete realisations} of {\it integrable} flows between exact CFTs.

A number of open questions remain to be addressed. Firstly, it would be interesting to determine the exact nature of the CFTs in the cases not done in the present work. Secondly, one could compute the anomalous dimensions of current operators, as well as that of primary operators along the lines of \cite{Georgiou:2015nka,Georgiou:2016iom,Georgiou:2016zyo,Georgiou:2017aei,c-function:2018,Georgiou:2017oly} starting with the two -coupling model of section 3. Furthermore, the exact $C$-function of the models could be calculated as was done in \cite{c-function:2018} for simpler cases. We expect that these 
computations will be technically quite challenging since the two coupling will both enter non-trivially in the various expressions as we have seen in the expressions for the $\b$-functions. 

Another direction would be to study the case where all four couplings  are in play by determining the RG equations and identifying their fixed points. Compared to the two coupling case, we expect an even richer structure of the RG equations to be unveiled. 
In addition, one could search for integrability in the four coupling case. 
Finally, although it seems a formidable task, one could try to embed our models to solutions of type-IIB or type-IIA supergravity.

\section*{Acknowledgments}

K. S. would like to thank the Theoretical Physics Department of CERN for hospitality and 
financial support during part of this research. The work of G.G. on this project has received funding from the Hellenic Foundation for Research and Innovation
(HFRI) and the General Secretariat for Research and Technology (GSRT), under grant
agreement No 234.

\end{document}